\documentclass[aps,pra,twocolumn,floatfix,superscriptaddress,%
showpacs]{revtex4-1}

\usepackage{amsmath,amssymb}
\usepackage{bm,graphicx,color}

\newcommand{\bk}{\mathbf{k}}
\newcommand{\bulk}{\mathrm{bulk}}
\newcommand{\bz}{\mathbf{z}}
\newcommand{\dn}{\downarrow}
\newcommand{\Ed}{\eps_d}
\newcommand{\Eds}{\eps_{d,\sigma}}
\newcommand{\EF}{\eps_F}
\newcommand{\eff}{\mathrm{eff}}
\newcommand{\Emin}{\tilde{\eps}_0}
\newcommand{\eps}{\epsilon}
\newcommand{\ER}{\eps_R}
\newcommand{\Ezero}{\eps_0}
\newcommand{\EZ}{\eps_{\hspace*{-.2ex}Z}}
\newcommand{\half}{\mbox{$\frac{1}{2}$}}
\newcommand{\hyb}{\mathrm{hyb}}
\newcommand{\imp}{\mathrm{imp}}
\newcommand{\J}{\mathbf{J}}
\newcommand{\Mimp}{M_{\imp}}

\newcommand{\pdag}{\phantom{\dag}}
\newcommand{\Rash}{\mathrm{Rashba}}
\newcommand{\rhoz}{\varrho_0}
\newcommand{\Sd}{\mathbf{S}_d}
\newcommand{\tc}{\tilde{c}}
\newcommand{\tEd}{\tilde{\eps}_d}
\newcommand{\tf}{\tilde{f}}
\newcommand{\trho}{\tilde{\rho}}
\newcommand{\tU}{\tilde{U}}
\newcommand{\tV}{\tilde{V}}
\newcommand{\up}{\uparrow}
\newcommand{\Zee}{\mathrm{Zeeman}}

\begin{document}

\title{Influence of Rashba spin-orbit coupling on the Kondo effect}

\date{\today}

\author{Arturo Wong}
\altaffiliation[Present address: ]{Centro de Nanociencas y Nanotecnolog\'{i}a,
Universidad Nacional Aut\'{o}noma de M\'{e}xico, Ensenada, Baja California,
22800, Mexico}
\affiliation{Department of Physics, University of Florida, P.O.\ Box 118440,
Gainesville, Florida 32611-8440, USA}
\author{Sergio E.\ Ulloa}
\author{Nancy Sandler}
\affiliation{Department of Physics and Astronomy, Nanoscale and Quantum
Phenomena Institute, Ohio University, Athens, Ohio 45701, USA}
\author{Kevin Ingersent}
\affiliation{Department of Physics, University of Florida, P.O.\ Box 118440,
Gainesville, Florida 32611-8440, USA}

\begin{abstract}
An Anderson model for a magnetic impurity in a two-dimensional electron gas
with bulk Rashba spin-orbit interaction is solved using the numerical
renormalization group under two different experimental scenarios. For a fixed
Fermi energy, the Kondo temperature $T_K$ varies weakly with Rashba coupling
$\alpha$, as reported previously. If instead the band filling is low and held
constant, increasing $\alpha$ can drive the system into a helical regime with
exponential enhancement of $T_K$. Under either scenario, thermodynamic
properties at low temperatures $T$ exhibit the same dependences on $T/T_K$ as
are found for $\alpha=0$. Unlike the conventional Kondo effect, however, the
impurity exhibits static spin correlations with conduction electrons of nonzero
orbital angular momentum about the impurity site.
We also consider a magnetic field that Zeeman splits the conduction band but
not the impurity level, an effective picture that arises under a proposed
route to access the helical regime in a driven system. The impurity
contribution to the system's ground-state angular momentum is found to be a
universal function of the ratio of the Zeeman energy to a temperature scale
that is not $T_K$ (as would be the case in a magnetic field that couples
directly to the impurity spin), but rather is proportional to $T_K$ divided by
the impurity hybridization width. This universal scaling is explained via a
perturbative treatment of field-induced changes in the electronic density of
states.
\end{abstract}


\maketitle

\section{Introduction}

The field of spintronics has primarily been driven by the idea of manipulating
spin states to create new, spin-based electronic devices \cite{Wolf:2001,
Zutic:2004}. Rashba spin-orbit (SO) coupling \cite{Winkler:2003} has been
proposed as a mechanism for spin-control, not only because of the possibility
of its external manipulation, but also because it is the physical origin of
spin-dependent phenomena such anisotropic magnetoresistance
\cite{Jungwirth:2006} and the spin-Hall effect \cite{Jungwirth:2012}.
Interest in SO interactions has also been motivated by the recent discovery of
large Rashba spin-splittings in Bi$_2$Se$_3$ topological insulators
\cite{King:2011,Zhu:2011}, where the Rashba parameter can be an order of
magnitude higher than in standard III-V semiconductors.

The study of Kondo correlations in the presence of SO interactions can be
traced back more than 40 years to experiments that seemed to demonstrate
suppression of the Kondo effect by Pt impurities \cite{Gainon-Heeger:1969}.
However, early theoretical studies of Anderson and Kondo models including SO
scattering from heavy nonmagnetic impurities reached opposing conclusions as to
whether SO interactions cut off the Kondo $\ln(T/T_K)$ term in the
resistivity \cite{Giovanni:1971,Everts:1972}. Subsequent magnetoresistance
measurements were interpreted as providing evidence for coexistence of SO
scattering and the Kondo effect \cite{Bergmann:1986}. A similar conclusion was
reached on the basis of time-reversal symmetry \cite{Meir-Wingreen:1994},
although this assertion has recently been challenged \cite{Eriksson:2012}.

In recent years, several theoretical works have investigated the effect of SO
interaction of the Rashba type on the Kondo temperature $T_K$. An analysis of
the Kondo model for a magnetic impurity in an otherwise clean two-dimensional
electron gas (2DEG) concluded that $T_K$ remains essentially unchanged by
Rashba coupling \cite{Malecki:2007}. A numerical renormalization-group (NRG)
study of an Anderson model describing the same physical situation similarly
predicted weak enhancement or depression of $T_K$, depending on the energy of
the impurity level relative to the Fermi energy \cite{Zitko-Bonca:2011}. A
variational treatment of the Kondo problem for arbitrary band dispersion and a
general SO coupling also found no significant change of $T_K$, although it was
claimed that the impurity is only partially screened \cite{Isaev:2012}. By
contrast, a mapping via a generalized Schrieffer-Wolff transformation
\cite{Schrieffer:1966} of an Anderson impurity in a two-dimensional host to an
effective Kondo model led to the prediction \cite{Zarea:2012} of an exponential
enhancement of $T_K$. In the specific context of adatoms on graphene, it was
shown \cite{Mastrogiuseppe:2014} that Kondo physics survives the presence of
bulk Rashba coupling, with a Kondo temperature that can change faster or slower
with tuning of the chemical potential than would be the case in the absence of
SO interaction (where the low-energy excitations are massless Dirac fermions).

This paper revisits the Kondo problem in the presence of Rashba SO interaction
from a different perspective. We focus on two different scenarios under which
the Rashba coupling in a 2DEG might be externally tuned: (1) an open electron
system with a Fermi level pinned to that of external reservoirs, and (2) an
isolated system with a constant band filling and a Fermi energy that varies
with the Rashba coupling.
The appropriate model for a magnetic impurity in a Rashba-coupled host is
mapped exactly to effective two-channel and one-channel Anderson models without
SO interaction but with conduction-band densities of states that are modified
to account for the Rashba coupling. For the specific case of quadratic band
dispersion (in the absence of SO interaction) and local impurity-band
hybridization, we use the NRG technique to solve the effective one-channel
model to calculate thermodynamic properties, from which we extract the Kondo
temperature.
For fixed Fermi energy, the case considered in previous work, we reproduce the
conclusions of Ref.\ \onlinecite{Zitko-Bonca:2011} that many-body screening of
the impurity is complete, thermodynamic properties have conventional Kondo
temperature dependences, and varying the Rashba coupling produces only modest
changes in the many-body scale $T_K$. For fixed band filling, by contrast,
increasing the Rashba coupling can drive the system into a helical regime with
an increase in the effective density of states at the Fermi level. In the
helical regime, thermodynamics retain their conventional nature, but with a
characteristic scale $T_K$ that is exponentially enhanced.

We also solve numerically the effective two-channel Anderson model, which
retains angular-momentum information that is discarded in the one-channel model.
Calculations of static angular-momentum correlations provide explicit
confirmation of the expectation from previous works \cite{Malecki:2007,
Zitko-Bonca:2011,Isaev:2012,Zarea:2012,Mastrogiuseppe:2014} that the SO
interaction induces an indirect coupling between the impurity and electrons of
nonzero orbital angular momentum.

Entry to the helical regime requires very strong Rashba couplings and/or low
carrier densities \cite{Winkler:2003}. There has been considerable recent
progress in fabrication of very low-density and clean two-dimensional hole gases
\cite{Huang:2014}, but here Coulomb interactions will likely replace disorder
as a barrier to reaching the helical regime. However, it has been suggested that
this regime may be accessed in a driven system by using circularly polarized
light to create an effective Zeeman field that opens a gap between the two
Rashba bands \cite{Ojanen-Kitagawa:2012}. (Similar ideas have been proposed for
engineering topological states in insulators \cite{Lindner:2011,Dora:2012}.)
Although the breaking of time-reversal symmetry is inimical to the Kondo effect,
this proposed experiment offers an opportunity to study Kondo physics in the
presence of an effective magnetic field that couples directly only to the bulk
electrons.

A real magnetic field would couple both to the spin of the bulk electrons and
to the impurity spin, with respective $g$ factors $g_b$ and $g_i$ that need
not be equal.
There have been numerous studies of Anderson and Kondo models in fields that
couple equally to the bulk and impurity spins ($g_b=g_i$) or only to the
impurity ($g_b=0$). Moreover, it is has been shown \cite{Vigman} how to map
between Kondo models having different pairs $(g_b,\,g_i)$ and $(g'_b,\,g'_i)$
so long as $g_i\ne 0$ and $g'_i\ne 0$. It is well understood \cite{Andrei} that
in such cases, Kondo correlations are destroyed once the conduction-band Zeeman
splitting $2\,\EZ$ becomes comparable \cite{units} to $T_K$. By contrast, the
proposal of Ref.\ \onlinecite{Ojanen-Kitagawa:2012} corresponds to a case
$g_i=0$ that has received little attention until now. Under these
circumstances, we find (through NRG solution of an effective one-channel
Anderson model) that the impurity contribution to the total angular momentum of
the system's ground state is a universal function, not of $\EZ/T_K$, but rather
of $f \,\Gamma\EZ / T_K D$, where $\Gamma$ is the hybridization width of the
impurity level, $D$ is a measure of the conduction-band width, and the
dimensionless quantity $f$ depends on other model parameters: the impurity level
energy and on-site Coulomb repulsion, as well as (in this particular
realization) the Rashba coupling. A perturbative treatment of field-induced
changes in the effective densities of states for electrons with different
components of the total angular momentum allows the scaling to be interpreted
in terms of an effective spin-splitting of the impurity level by an energy
proportional to $\Gamma\EZ / D$. A similar picture should hold for any
realization of the Anderson impurity model with $g_i = 0$,  as seems likely to
be achievable in lateral quantum dots \cite{Allison:2014}.

The remainder of the paper is organized as follows. Section \ref{sec:model}
describes the Anderson model for a magnetic impurity in a two-dimensional host
with bulk Rashba SO interaction, and outlines the mapping of the problem to
a one-channel Anderson model with a hybridization function that depends on
energy and, in the presence of a bulk Zeeman splitting as proposed in Ref.\
\onlinecite{Ojanen-Kitagawa:2012}, also on the component of the electron's
total angular momentum parallel to the Zeeman field. Explicit expressions for
the hybridization function are provided for cases where the band dispersion in
the absence of Rashba coupling is purely quadratic. Section \ref{sec:results}
presents numerical results for such cases, focusing on the effect of the
Rashba coupling on the Kondo temperature $T_K$ and on static angular-momentum
correlations, as well as the variation of the impurity polarization with bulk
Zeeman field.
We summarize our results in Sec.\ \ref{sec:summary}. Appendix \ref{app:pert}
describes a perturbative method used to analyze the effects of Rashba coupling
(in Sec.\ \ref{subsubsec:TK}) and of a bulk Zeeman splitting (in Sec.\
\ref{subsec:Zfield}). Details of the calculation of angular-momentum
correlations appear in Appendix \ref{app:spin}.

\section{Model and Preliminary Analysis}
\label{sec:model}

\subsection{Anderson model with Rashba coupling}
\label{subsec:model}

We consider an Anderson impurity in a two-dimensional electron gas in the
presence of Rashba SO coupling, modeled by the Hamiltonian
\cite{Zitko-Bonca:2011,Zarea:2012} 
\begin{equation}
\label{H_orig}
H=H_{\bulk}+H_{\imp}+H_{\hyb}.
\end{equation}
Here, $H_{\bulk}=H_0+H_{\Rash}$, where
\begin{equation}
\label{H_band}
H_0 = \sum_{\bk,\sigma} \eps({\bk}) \, c_{\bk,\sigma}^{\dag}
  c_{\bk,\sigma}^{\pdag}
\end{equation}
describes the conduction band in the absence of SO interaction, with operator 
$c_{\bk,\sigma}$ destroying a band electron of two-dimensional wave vector
$\bk=k_x \hat{\mathbf{x}}+k_y \hat{\mathbf{y}}$, spin $z$ component
$\sigma=\pm 1/2$ (or $\up,\,\dn$), and energy $\eps({\bk})$. The second
term in $H_{\bulk}$ represents the effect of the Rashba SO interaction
$\lambda_R \, \hat{\bz} \cdot \bm{\sigma} \times \bk$, where $\bm{\sigma}/2$
is the electron spin operator:
\begin{equation}
\label{H_Rash}
H_{\Rash} = i \lambda_R \sum_{\bk} k \, e^{-i\phi_k}
  c_{\bk,\up}^{\dag} c_{\bk,\dn}^{\pdag} + \text{H.c.} ,
\end{equation}
where $k=|\bk|$, $\phi_k=\mathrm{atan} \, (k_y/k_x)$ are the polar
components of $\bk$, and $\lambda_R$ is the SO coupling (assumed in our
analysis to be non-negative).

In isolation, the nondegenerate impurity level is described by
\begin{equation}
\label{H_imp}
H_{\imp} = (\Ed+\mu) \bigl( d_{\up}^{\dag} d_{\up}^{\pdag}
  + d_{\dn}^{\dag} d_{\dn}^{\pdag} \bigr)
  + U d_{\up}^{\dag} d_{\up}d_{\dn}^{\dag} d_{\dn} ,
\end{equation}
where $d_{\sigma}$ destroys an electron with spin $z$ component $\sigma$ and
energy $\Ed$ relative to the chemical potential $\mu$, and $U$ is the on-site
Coulomb repulsion. The impurity state is assumed to exhibit axial symmetry
about $\hat{\mathbf{z}}$.

The last term in Eq.\ \eqref{H_orig}, representing tunneling of electrons
between the impurity and the bulk, is
\begin{equation}
\label{H_hyb}
H_{\hyb} = \frac{1}{\sqrt{N_c}} \sum_{\bk,\sigma} V(\bk)
  \bigl( c_{\bk,\sigma}^{\dag} d_{\sigma}^{\pdag} + \text{H.c.} \bigr) ,
\end{equation}
where $N_c$ is the number of unit cells in the host (and hence the number of
distinct $\bk$ values in the first Brillouin zone) and the hybridization matrix
element $V(\bk)$ can be taken to be real and non-negative. We note that
although the orbital motion of the conduction electrons is constrained to two
dimensions, all spin vectors are fully three-dimensional.

For simplicity, we consider a jellium host such that the band dispersion and
the hybridization matrix element are isotropic in $\bk$ space, i.e.,
$\eps({\bk})=\eps(k)$ and $V(\bk)=V(k)$.

\subsection{Mapping to a two-channel Anderson model}
\label{subsec:mapping1}

This section lays out an exact transformation of the Hamiltonian
\eqref{H_orig} into the form of an effective two-channel Anderson model for
a magnetic impurity hybridizing with two bands in which the SO interaction has
been subsumed into a modification of the density of states. The mapping
generalizes the one presented in Ref.\ \onlinecite{Zitko-Bonca:2011} to allow
for arbitrary forms of $\eps(k)$ and $V(k)$.

We take the thermodynamic limit in the standard manner by letting the unit-cell
number $N_c\to\infty$ and the system area $A\to\infty$ in such a way that
$A/N_c\to A_c$, a finite unit-cell area.
Each summation $\sum_{\bk} f(\bk)$ over a discrete wave vector $\bk$ can be
replaced by an integral $(A_c/4\pi^2) \int \! d^2 \bk \, f(\bk)$.

In the absence of SO coupling, it is natural to adopt a basis of states having a
definite $z$ component of the orbital angular momentum about the impurity site.
The transformation \cite{Malecki:2007,shift-note}
\begin{equation}
\label{transTAM}
c_{\bk,\sigma}^{\pdag} \to \sum_{m=-\infty}^{\infty}
  \sqrt{\frac{2\pi}{A_c k}} \, e^{im(\phi_k-\pi/2)} \, c_{k,m,\sigma} ,
\end{equation}
where
$\{c_{k,m,\sigma}^{\pdag},\, c_{k',m',\sigma'}^{\dag}\} =
  \delta(k-k')\,\delta_{m,m'}\delta_{\sigma,\sigma'}$,
allows one to rewrite Eq.\ \eqref{H_band} in the diagonal form
\begin{equation}
H_0 = \sum_{m,\sigma} \: \int \! dk \; \eps(k) \:
   c_{k,m,\sigma}^{\dag} \, c_{k,m,\sigma}^{\pdag},
\end{equation}
while the hybridization term becomes
\begin{equation}
\label{H_hyb:v2}
H_{\hyb} = \sqrt{\frac{A_c}{2\pi}} \int \! dk \; \sqrt{k} \, V(k) \,
  \bigl( c_{k,0,\sigma}^{\dag} d_{\sigma}^{\pdag} + \text{H.c.} \bigr) ,
\end{equation}
in which the impurity couples only to the $m=0$ mode.

The Rashba Hamiltonian term, which becomes
\begin{equation}
H_{\Rash} = \lambda_R \int \! dk \; k \sum_m \:
   c_{k,m,\up}^{\dag} \, c_{k,m+1,\dn}^{\pdag} + \text{H.c.},
\end{equation}
is not diagonal in the $(k,m,\sigma)$ basis because Rashba SO interaction
couples spin and orbital degrees of freedom. However, since $H_{\Rash}$
mixes only pairs of states $(k,m,\,\up)$ and $(k,m+1,\,\dn)$, $H_{\bulk}$
conserves $\tau = m + \sigma$, the $z$ component of total (orbital plus spin)
angular momentum. The bulk Hamiltonian also commutes with the helicity
operator
\begin{equation}
\hat{h}
  = \hat{\bz} \cdot \bm{\sigma} \times \hat{\bk}
  = \int \! dk \; \sum_m c_{k,m,\up}^{\dag} \; c_{k,m+1,\dn}^{\pdag}
    + \text{H.c.}
\end{equation} 
It is therefore convenient to perform a canonical transformation to a
new complete basis of fermionic operators
\begin{equation}
\label{a_kht}
\tc_{k,h,\tau}=\frac{1}{\sqrt{2}}
 \bigl( h^{\tau-1/2} \, c_{k,\tau-1/2,\up}
 + h^{\tau+1/2} \, c_{k,\tau+1/2,\dn} \bigr) ,
\end{equation}
each of which annihilates an electron in a state of well-defined
$\tau = \pm 1/2, \, \pm 3/2, \ldots$ and definite helicity
$h=\pm 1$ (abbreviated $h=\pm$ at certain points below).

This transformation diagonalizes the bulk Hamiltonian, yielding
\begin{equation}
\label{H_bulk:diag}
H_{\bulk} = \sum_{h,\tau} \int \! dk \;
  \eps_h(k) \: \tc_{k,h,\tau}^{\dag} \, \tc_{k,h,\tau}^{\pdag}
\end{equation}
with a helicity-dependent (but total-angular-momentum-independent) dispersion
\begin{equation}
\label{epsilon_h}
\eps_h(k) = \eps(k) + h \lambda_R k .
\end{equation}

The two operators $c_{k0\sigma}$ to which the impurity couples in Eq.\
\eqref{H_hyb:v2} can be represented in terms of four operators
$\tc_{k,h,\tau}$, namely those
with $h=\pm$ and $\tau = \pm 1/2$. Since these four operators also involve
$c_{k,-1,\up}$ and $c_{k,1,\dn}$, one sees that the Rashba SO interaction
creates an indirect coupling of the impurity to conduction electrons 
with nonzero orbital angular momentum \cite{Malecki:2007,Zarea:2012}. 

One can drop the uninteresting contribution to $H_{\bulk}$ from electrons
having total angular momentum $z$ component $|\tau| > 1/2$, thereby reducing
Eq.\ \eqref{H_orig} to a two-channel Anderson Hamiltonian with the helicity
$h$ acting as a channel index:
\begin{multline}
\label{H_2chan:k-space}
H = \sum_{h,\tau} \int \! dk \; \eps_h(k) \: \tc_{k,h,\tau}^{\dag}
    \tc_{k,h,\tau}^{\pdag} + H_{\imp} \\
    + \sqrt{\frac{A_c}{4\pi}} \sum_{h,\tau} \int \! dk \; \sqrt{k} \, V(k) \,
    \bigl(\tc_{k,h,\tau}^{\dag} d_{\tau}^{\pdag}+\text{H.c.}\bigr) .
\end{multline}
Due to the difference $\eps_+(k)-\eps_-(k) = 2\lambda_R k$, the two
helicities enter Eq.\ \eqref{H_2chan:k-space} in an inequivalent manner; in
particular, they have different Fermi wave vectors. It is also important to
bear in mind that the index $\tau=\pm 1/2$ (or $\up$, $\dn$)
labels the $z$ component of the total angular momentum, although this reduces
to the $z$ component of spin for the impurity operators $d_{\tau}$.

Equation \eqref{H_2chan:k-space} can be transformed to an energy representation
by defining
\begin{equation}
\label{a_ehjs}
\tc_{\eps,h,j,\tau}
  = |\eps'_h(k_j)|^{-1/2} \; \tc_{k_j,h,\tau} \,,
\end{equation}
where $\eps'_h = d\eps_h/dk$ and $k_j(\eps,h)$, $j=1,\,\ldots,\,n_h(\eps)$ are
the numerically distinct roots of the equation $\eps_h(k_j)=\eps$. If
$\eps_h(k)$ is a monotonically increasing function of $k$, as would be the case
for free fermions in the absence of SO interaction, then $n_h(\eps)=0$ for
$\eps<\eps_h(0)$ and $n_h(\eps)=1$ for $\eps\ge\eps_h(0)$. However, as
discussed in greater detail in Sec.\ \ref{subsec:concrete}, the presence of
Rashba SO interaction creates an energy range within which $n_-(\eps)=2$.

The operators defined in Eq.\ \eqref{a_ehjs} obey the canonical anticommutation
relations
\begin{equation}
\bigl\{\tc_{\eps,h,j,\tau}^{\pdag}, \, \tc_{\eps',h',j',\tau'}^{\dag}\bigr\}
   = \delta(\eps-\eps') \, \delta_{h,h'} \, \delta_{j,j'} \,
   \delta_{\tau,\tau'}
\end{equation}
and allow Eq.\ \eqref{H_2chan:k-space} to be rewritten
\begin{multline}
\label{H_energy}
H = \sum_{h,\tau} \int \! d\eps \; \eps \sum_{j=1}^{n_h(\eps)}
    \tc_{\eps,h,j,\tau}^{\dag} \, \tc_{\eps,h,j,\tau}^{\pdag}
  + H_{\imp} \\
  + \sum_{h,\tau} \int \! d\eps \sum_{j=1}^{n_h(\eps)}
    \sqrt{\Gamma_{h,j}(\eps)/\pi} \: \bigl( \tc_{\eps,h,j,\tau}^{\dag} \,
    d_{\tau}^{\pdag} + \text{H.c.} \bigr) ,
\end{multline}
where
\begin{equation}
\Gamma_{h,j}(\eps)
  = \frac{A_c \, k_j}{4|\eps'_h(k_j)|} \, V(k_j)^2
\end{equation}
is the contribution to the helicity-$h$ hybridization function at energy
$\eps$ that arises from wave vector $k=k_j(\eps,h)$.

Equation \eqref{H_energy} is an exact restatement of Eq.\
\eqref{H_2chan:k-space}, and allows full recovery of dependences on the radial
coordinate measured from the impurity site, as obtained via Fourier
transformation with respect to $k$. Further transformations of the Hamiltonian
described in Sec.\ \ref{subsec:mapping2} below serve to simplify the calculation
of certain thermodynamic properties, but necessarily entail loss of information
about radial or angular momentum degrees of freedom that could be inferred from
a complete solution of Eq.\ \eqref{H_energy}.

\subsection{Further reduction of the model}
\label{subsec:mapping2}

One simplification of Eq.\ \eqref{H_energy} arises from noting that for
energies $\eps$ where $n_h(\eps) > 1$, the impurity couples to a single
linear combination of the operators $\tc_{\eps hj\tau}$,
$j=1,\,2,\,\ldots,\,n_h(\eps)$. Defining
\begin{equation}
\sqrt{\Gamma_h(\eps)} \; \tc_{\eps,h,\tau}
  = \sum_{j=1}^{n_h(\eps)} \sqrt{\Gamma_{h,j}(\eps)}
    \; \tc_{\eps,h,j,\tau}
\end{equation}
with a helicity-$h$ hybridization function
\begin{equation}
\Gamma_h(\eps)
 = \sum_{j=1}^{n_h(\eps)} \Gamma_{h,j}(\eps)
\end{equation}
allows one to write
\begin{multline}
\label{H_2chan}
H = \sum_{h,\tau} \int \! d\eps \; \eps \:
    \tc_{\eps,h,\tau}^{\dag} \, \tc_{\eps,h,\tau}^{\pdag}
  + H_{\imp} \\
  + \sum_{h,\tau} \int \! d\eps \:
    \sqrt{\Gamma_h(\eps)/\pi}
    \: \bigl( \tc_{\eps,h,\tau}^{\dag} \, d_{\tau}^{\pdag}
    + \text{H.c.} \bigr) ,
\end{multline}
from which have been dropped diagonal terms involving $n_h(\eps)-1$
linear combinations of the operators $\tc_{\eps,h,j,\tau}$ that are
orthogonal to $\tc_{\eps,h,\tau}$. The mapping from $n_h(\eps) > 1$
operators $\tc_{\eps,h,j,\tau}$ to a single $\tc_{\eps,h,\tau}$ involves
loss of radial information since the latter operator cannot be associated
with any single wave vector $k$.

Another simplification can be made by combining the $h=+$ and $h=-$ states of
the same energy that couple to the impurity. Defining
\begin{equation}
\sqrt{\Gamma(\eps)} \; \tc_{\eps,\tau}
  = \sqrt{\Gamma_+(\eps)} \; \tc_{\eps,+,\tau}
  + \sqrt{\Gamma_-(\eps)} \; \tc_{\eps,-,\tau}
\end{equation}
with a total hybridization function
\begin{equation}
\Gamma(\eps) = \Gamma_+(\eps) + \Gamma_-(\eps),
\end{equation}
one can again discard decoupled degrees of freedom [here, associated with
$\sqrt{\Gamma_-(\eps)} \; \tc_{\eps,+,\tau} - \sqrt{\Gamma_+(\eps)}
\; \tc_{\eps,-,\tau}$] to arrive at an effective one-impurity Anderson model
\begin{multline}
\label{H_1chan}
H = \sum_{\tau} \int \! d\eps \; \eps \:
    \tc_{\eps,\tau}^{\dag} \, \tc_{\eps,\tau}^{\pdag}
  + H_{\imp} \\
  + \sum_{\tau} \int \! d\eps \:
    \sqrt{\Gamma(\eps)/\pi}
    \: \bigl( \tc_{\eps,\tau}^{\dag} \, d_{\tau}^{\pdag}
    + \text{H.c.} \bigr) .
\end{multline}

That the original model in Eq.\ \eqref{H_orig} can be reduced to a one-channel
Anderson impurity model was shown previously in Ref.\
\onlinecite{Zitko-Bonca:2011} for the specific case of a quadratic $\eps(k)$
and a local ($k$-independent) $V(k)$. However, the derivation above makes clear
that the price paid for going from Eq.\ \eqref{H_2chan} to Eq.\ \eqref{H_1chan}
is the loss of the ability to distinguish between the spin and orbital angular
momenta within a bulk state of given $z$ component of the total angular momentum.
For this reason, Sec.\ \ref{sec:results} presents not only impurity properties
calculated from Eq.\ \eqref{H_1chan}, but also impurity-bulk angular-momentum
correlations obtained via numerical solution of Eq.\ \eqref{H_2chan}.

In the absence of the impurity, the bands entering Eqs.\ \eqref{H_2chan}
and \eqref{H_1chan} would be filled at temperature $T=0$ up to a chemical
potential $\mu=\EF$. (As will be emphasized in Secs.\ \ref{subsec:concrete} and
\ref{sec:results}, the Fermi energy $\EF$ may or may not take the same value as
for $\alpha=0$, depending on the experimental setup being described.) In cases
where $\Gamma(-\Ed)\ll-\Ed$ and $\Gamma(U+\Ed)\ll U+\Ed$, it is appropriate to
apply a generalized Schrieffer-Wolff transformation \cite{Schrieffer:1966} to
map the two- and one-channel Anderson models to two- and one-channel Kondo
models, respectively. In general, the resulting Kondo models \cite{Zarea:2012}
will not be the same as those obtained by starting with a Kondo Hamiltonian
for a magnetic impurity in a 2DEG and then incorporating a bulk Rashba SO
interaction.

\subsection{Model with bulk Zeeman field}

In a recent paper, Ojanen and Kitagawa \cite{Ojanen-Kitagawa:2012} proposed to
realize a helical system through irradiation of a two-dimensional electron gas
containing Rashba SO interaction by light in the THz frequency range. After
time-averaging over a period of the electromagnetic radiation, the bulk
electrons experience an effective Zeeman coupling of tunable strength
$\EZ = (\lambda_R e E_0)^2/\Omega^3$, where $E_0$ and $\Omega/2\pi$
are the magnitude and frequency of the applied electric field.
Provided that the characteristic rate $k_B T_K/2\pi$ \cite{units} of spin
flips involved in Kondo screening is much slower than $\Omega/2\pi$, the
impurity will effectively interact with the time-averaged band structure. This
regime spans $T_K\ll 50$\,K for $\Omega/2\pi=1$\,THz and $T_K\ll 500$\,K for
$\Omega/2\pi=10$\,THz, conditions that will be readily satisfied in most
experiments. We also note that for the values of $E_0$ envisioned in
Ref.\ \onlinecite{Ojanen-Kitagawa:2012}, the magnetic component of the
circularly polarized light is so small as to have negligible effect. 

With this motivation, we consider the Hamiltonian \eqref{H_orig} augmented by
a term
\begin{equation}
\label{H_Z}
H_{\Zee}
  = 2 \EZ\sum_{\bk,\sigma}\sigma \, c_{\bk,\sigma}^{\dag} c_{\bk,\sigma}^{\pdag},
\end{equation}
where $\sigma=\pm 1/2$ or $\up$, $\dn$, depending on the context,
and we assume below that $\EZ\ge 0$.
The model can again be mapped to effective two-channel and one-channel
Anderson models via a sequence of steps along the lines laid out in Secs.\
\ref{subsec:mapping1} and \ref{subsec:mapping2}.

In order to diagonalize
$H_{\bulk}=H_0+H_{\Rash}+H_{\Zee}$,
the operator transformation in Eq.\ \eqref{a_kht} must be generalized to
\begin{eqnarray}
\label{a_kht:v2}
\tc_{k,h,\tau} &=& \frac{1}{\sqrt{2}}
  \bigl [ h^{\tau-1/2} \, \beta_h(k) \, c_{k,\tau-1/2,\up} \\ \nonumber
        &+& h^{\tau+1/2} \, \beta_{-h}(k) \, c_{k,\tau+1/2,\dn}
  \bigr] ,
\end{eqnarray}
where
\begin{equation}
\label{beta:def}
\beta_{\pm 1}(k) = \sqrt{1 \pm
  \frac{\EZ}{(\lambda_R^2 k^2 + \EZ^2)^{1/2}}} \, .
\end{equation}

This yields Eq.\ \eqref{H_bulk:diag} with a helicity-dependent dispersion
\begin{equation}
\label{epsilon_h:v2}
\eps_h(k) = \eps(k) + h \sqrt{\lambda_R^2 k^2 + \EZ^2}
\end{equation}
that features a gap $2\,\EZ$ at $k=0$, while the
impurity-bulk hybridization becomes
\begin{equation}
\label{H_hyb:v3}
H_{\hyb}
  = \sqrt{\frac{A_c}{4\pi}} \sum_{h,\tau} \int \! dk \; \sqrt{k} \, V(k) \,
    \beta_{2\tau h}(k) \,
    \bigl( \tc_{k,h,\tau}^{\dag} \, d_{\tau}^{\pdag} + \text{H.c.} \bigr) ,
\end{equation}
where the value of the product $2\tau h = \pm 1$ selects between the two
functions $\beta_{\pm 1}(k)$ defined in Eq.\ \eqref{beta:def}.
After further transformation to an energy representation, the problem maps to
a generalized two-channel Anderson model
\begin{multline}
\label{H_2chan:Z}
H = \sum_{h,\tau} \int \! d\eps \; \eps \:
    \tc_{\eps,h,\tau}^{\dag} \, \tc_{\eps,h,\tau}^{\pdag}
  + H_{\imp} \\
  + \sum_{h,\tau} \int \! d\eps \:
    \sqrt{\Gamma_{h,\tau}(\eps)/\pi}
    \: \bigl( \tc_{\eps,h,\tau}^{\dag} \, d_{\tau}^{\pdag}
    + \text{H.c.} \bigr) ,
\end{multline}
containing a helicity- and angular-momentum-dependent hybridization function
\begin{equation}
\Gamma_{h,\tau}(\eps)
  = \sum_{j=1}^{n_h(\eps)} \frac{A_c \, k_j}{4|\eps'_h(k_j)|} \,
    \bigl[ \beta_{2\tau h}(k_j) \, V(k_j) \bigr]^2 .
\end{equation}
Here, $k_j(\eps,h)$, $j=1,\,\ldots,\,n_h(\eps)$ are the distinct
roots of the equation $\eps_h(k_j)=\eps$ for the gapped dispersion in
Eq.\ \eqref{epsilon_h:v2}.

As before, the two-channel Anderson model can be mapped into an effective
one-channel model. It is straightforward to show that the impurity couples
only to the linear combination of operators defined via
\begin{equation}
\sqrt{\Gamma_{\tau}(\eps)} \; \tc_{\eps,\tau}
  = \sqrt{\Gamma_{+,\tau}(\eps)} \; \tc_{\eps,+,\tau}
  + \sqrt{\Gamma_{-,\tau}(\eps)} \; \tc_{\eps,-,\tau}
\end{equation}
with
\begin{equation}
\Gamma_{\tau}(\eps)
  = \Gamma_{+,\tau}(\eps) + \Gamma_{-,\tau}(\eps) ,
\end{equation}
leading to a Hamiltonian
\begin{multline}
\label{H_1chan:Z}
H = \sum_{\tau} \int \! d\eps \; \eps \:
    \tc_{\eps,\tau}^{\dag} \, \tc_{\eps,\tau}^{\pdag}
  + H_{\imp} \\
  + \sum_{\tau} \int \! d\eps \:
    \sqrt{\Gamma_{\tau}(\eps)/\pi}
    \: \bigl( \tc_{\eps,\tau}^{\dag} \, d_{\tau}^{\pdag}
    + \text{H.c.} \bigr) .
\end{multline}
Comparison with Eq.\ \eqref{H_1chan} shows that the effect the Zeeman
field is subsumed into an angular-momentum dependence of the hybridization
function.

\subsection{Local hybridization and quadratic band dispersion}
\label{subsec:concrete}

The band dispersion $\eps(k)$, the hybridization matrix element $V(k)$, and
the Zeeman energy $\EZ$ enter Eqs.\ \eqref{H_2chan:Z} and \eqref{H_1chan:Z}
only in combination through the hybridization functions
$\Gamma_{h,\tau}(\eps)$, which reduce to $\Gamma_h(\eps)$ for $\EZ=0$.
Henceforth, we will assume that the hybridization is local, i.e., $V(k)=V$,
in which case each hybridization function can be written as an
energy-independent prefactor $\pi V^2$ times an appropriately resolved density
of states per unit cell. For example,
\begin{equation}
\label{constant_V}
\Gamma_{h,\tau}(\eps) = \pi \rho_{h,\tau}(\eps) \, V^2
\end{equation}
where $\rho_{h,\tau}(\eps)$ is the density of states per unit cell for
helicity-$h$ and total angular momentum $z$ component $\tau$.

We also specialize to cases in which the band dispersion in the absence
of SO interaction takes the purely parabolic form \cite{Zitko-Bonca:2011,units}
$\eps(k)=\Ezero+k^2/2m^*$, where $m^*$ is the effective mass and
$\Ezero\le 0$ is the position of the bottom of the band relative to the Fermi
energy $\EF=0$. This dispersion yields the density of states (per unit cell,
per spin orientation)
\begin{equation}
\label{rho_band}
\rho_0(\eps)
  = \frac{A_c \, k}{2\pi |d\eps/dk|}
  = \rhoz \: \Theta(\eps-\Ezero) \: \Theta(D-\eps)
\end{equation}
where $\rhoz = A_c m^*/(2\pi)$ and
$D \equiv \eps(k_{\mathrm{max}}) = \Ezero+\rhoz^{-1}$ is an upper cutoff
introduced to enforce
$\int_{-\infty}^{\infty} \rho(\eps) \, d\eps = 1$.
We consider situations where the band is less than half-filled (i.e.,
$|\Ezero|<D$) and take $D$ to be the fundamental energy scale in the problem.

\begin{figure}[tb]
\centerline{\includegraphics[width=0.95\columnwidth]{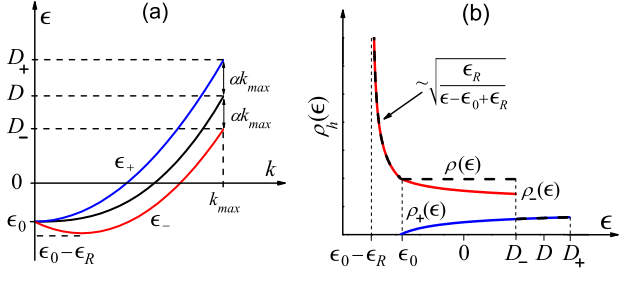}}
\caption{\label{fig:h-modes} (Color online)
Schematic plots of (a) the dispersion relations $\eps_h(k)$ and (b) the
densities of states per helicity channel $\rho_h(\eps)$, in the presence of
Rashba SO interaction. The middle curve in (a) represents the dispersion
$\eps(k)$ in the absence of Rashba interaction. In (b), the combined density
of states $\rho(\eps)$ (dashed line) is constant and equal to its no-Rashba
value $\rho_0(\eps)$ throughout the energy range $\Ezero<\eps<D_-$.}
\end{figure}

When Rashba SO interaction is taken into account, the helicity-$h$ dispersion
defined in Eq.\ \eqref{epsilon_h} can be expressed as
\begin{equation}
\label{epsilon_h:quad}
\eps_h(k) = \Emin + \frac{(k + h k_R)^2}{2m^*}, 
\end{equation}
where it is convenient to define a Rashba wave vector
$k_R = m^*\lambda_R$ and a Rashba energy
$\ER = k_R^2/2m^* = m^*\lambda_R^2/2 = \lambda_R k_R / 2$, such that
$\Emin = \Ezero - \ER$ is the energy at a parabolic minimum in $\eps_-(k)$
located at $k = k_R$. These dispersions, plotted schematically in Fig.\
\ref{fig:h-modes}(a), yield helicity-resolved densities of states (per unit
cell, per total angular momentum $z$ component)
\begin{equation}
\label{rho_h}
\rho_h(\eps) = \begin{cases}
  \: \rhoz \, \dfrac{\ER}{\sqrt{\ER(\eps-\Emin)}} \: \delta_{h,-}
    & \text{for } \Emin < \eps < \Ezero, \\[4ex]
  \dfrac{\rhoz}{2} \biggl[ 1
   - h \, \dfrac{\ER}{\sqrt{\ER(\eps-\Emin)}} \biggr]
    & \text{for } \Ezero < \eps < D_h , \\[3ex]
  0 & \text{otherwise.}
\end{cases}
\end{equation}
The upper cutoff of the helicity-$h$ band has shifted from $D$ to
$D_h = \eps_h(k_{\mathrm{max}}) = D + 2 h \sqrt{\ER(D-\Ezero)}$ such that
$\int_{-\infty}^{\infty} \rho_h(\eps) \, d\eps= \half$. The densities
of states $\rho_h(\eps)$ are plotted schematically in Fig.\
\ref{fig:h-modes}(b). When compared with $\half\rho_0(\eps)$
[to which $\rho_+(\eps)$ and $\rho_-(\eps)$ reduce for $\ER = 0$], the
most striking features are
(i) the shift of $h=+$ states to higher energies $D<\eps<D_+$, resulting
in a monotonic depression of $\rho_+(\eps)$ to zero as $\eps\to 0^+$,
and (ii) the shift of $h=-$ states from $D_-<\eps<D$ to lower energies, and
particularly the $1/\sqrt{\eps-\Emin}$ variation of $\rho_-(\eps)$ over
the range $\Emin<\eps<\Ezero$. The van Hove singularity in $\rho_-(\eps)$
at $\eps=\Emin$ arises from the parabolic minimum in $\eps_-(k)$ at
$k=k_R$.

Figure \ref{fig:h-modes}(b) also shows (dashed line) the density of states
$\rho(\eps)=\rho_+(\eps)+\rho_-(\eps)$ for the effective
one-channel Anderson problem defined in Eq.\ \eqref{H_1chan}. This function
is identical to its counterpart in the absence of Rashba interaction over a
wide energy window $0 \le \eps \le D_-$, with $\rho(\eps)$ differing
from $\rho_0$ only in the redistribution of weight around the
upper band edge ($\eps>D_-$), with part of that weight being transferred
into the low-energy upturn spanning $\Emin < \eps < \Ezero$.

In Sec.\ \ref{sec:results}, we examine the effect of increasing the Rashba
energy $\ER$, as might be achieved experimentally by increasing the strength of
an electric field applied perpendicularly to the two-dimensional electron
gas. We consider two scenarios:

\begin{figure}[tb]
\centerline{\includegraphics[width=0.95\columnwidth]{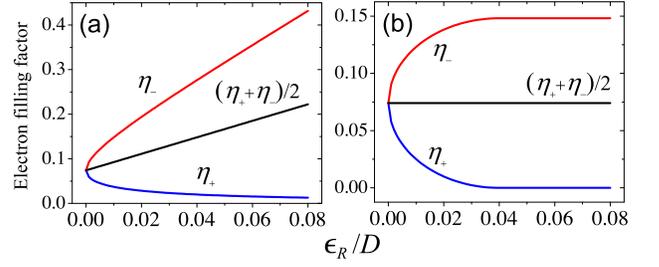}}
\caption{\label{fig:filling} (Color online)
Helicity-resolved conduction-band filling fractions $\eta_{\pm}$ and overall
filling fraction $\eta=(\eta_+ + \eta_-)/2$, plotted as functions of Rashba
energy $\ER$ for $\Ezero=-0.08D$ and
(a) fixed Fermi energy $\EF=0$,
(b) constant filling fraction $\eta=-\Ezero/(D-\Ezero)\approx 0.074$.}
\end{figure}

\noindent
(1) The Fermi energy is fixed at $\EF = 0$, as would be the case if the system
were maintained in equilibrium with a reservoir of electrons at fixed chemical
potential. As illustrated in Fig.\ \ref{fig:filling}(a), the filling fraction of
each helicity band, $\eta_h=2\int_{-\infty}^{\EF} \rho_h(\eps)\,d\eps$,
changes with the Rashba energy in such a way that the overall filling fraction
increases linearly with $\ER$:
\begin{equation}
\eta = \half (\eta_+ + \eta_-)
     = \frac{2\,\ER-\Ezero}{D-\Ezero} .
\end{equation}

\noindent
(2) The overall band filling is held constant, as would occur if the system were
isolated from any external source of electrons. In this case, as illustrated in
Fig.\ \ref{fig:filling}(b), varying the Rashba coupling still leads to changes
in $\eta_+$ and $\eta_-$, but it does not alter their mean $\eta=-\Ezero/(D-\Ezero)$.
This comes about because with increasing $\ER$, the Fermi energy decreases
according to
\begin{equation}
\label{EFconstN}
\EF(\ER)=\begin{cases}
  -2\,\ER & \ER<|\Ezero|/2, \\[1ex]
  \Ezero-\ER+\Ezero^2/4\ER & \ER\ge|\Ezero|/2.
\end{cases}
\end{equation}   
For $\ER> |\Ezero|/2$, the Fermi energy lies below $\Ezero$, with two
important consequences. First, the occupied bulk states all have $h=-$,
resulting in the formation of an unconventional, helical metal. Second, the
total density of states at the Fermi level is $\rho(\EF)=2\rhoz \ER/|\Ezero|$,
which is enhanced over its value $\rhoz$ for $\ER=0$. This situation is unlikely
to be realized in standard III-V semiconductor heterostructures
\cite{Winkler:2003}, since it would require very high Rashba couplings or very low
carrier densities that will be affected by disorder or Coulomb interactions.
However, driven systems may allow investigation of this interesting regime
provided that one takes into account the effective Zeeman splitting of the
helicity-resolved bands \cite{Ojanen-Kitagawa:2012} .

\begin{figure}[tb]
\centerline{\includegraphics[width=\columnwidth]{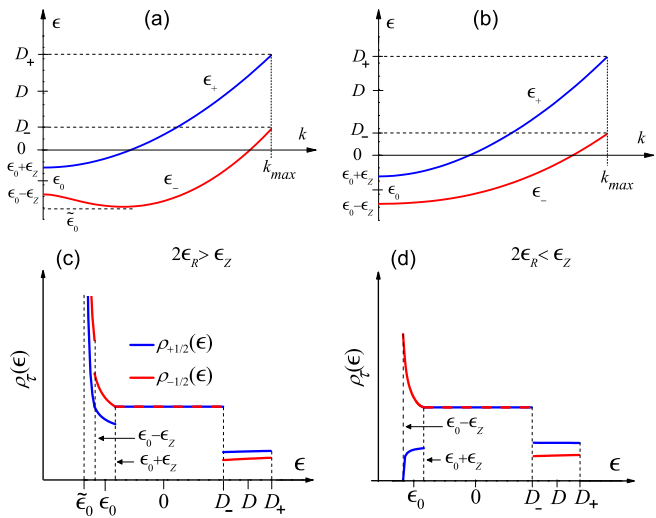}}
\caption{\label{fig:h+tau-modes} (Color online)
Schematic plots of the effective angular-momentum-resolved densities of
states $\rho_{\tau}(\eps)$ obtained from Eq.\ \eqref{rho_h,tau} for
(a) weak Zeeman splitting $\EZ<2\,\ER$,
(b) strong Zeeman splitting $\EZ>2\,\ER$.
In both panels, $\rho_{\tau}(\eps)$ is constant and equal to its no-Rashba
value $\rho_0(\eps)$ throughout the energy range $\Ezero+\EZ<\eps<D_-$.}
\end{figure}

In the presence of a bulk Zeeman field as well as Rashba interaction, the
helicity-$h$ dispersion in Eq.\ \eqref{epsilon_h:v2} can be expressed as
\begin{equation}
\label{epsilon_h:v2:quad}
\eps_h(k) = \Emin + \frac{(\sqrt{k^2 + k_Z^2} + h k_R)^2}{2m^*}, 
\end{equation}
where $k_Z=\EZ/\lambda_R$ and we have redefined
\begin{equation}
\Emin = \Ezero - \ER - \frac{\EZ^2}{4 \ER}
  = \Ezero -\EZ - \frac{(2 \ER -\EZ)^2}{4 \ER} .
\end{equation}
Equation \eqref{epsilon_h:v2:quad} implies that $\eps_+(k)$ rises
monotonically with increasing $k$ from $\eps_+(0) = \Ezero +\EZ$. If $k_Z < k_R$,
which is equivalent to the condition $\EZ < 2 \ER$, the $h=-$ dispersion has a
parabolic minimum at a nonzero wave vector $k=\sqrt{k_R^2-k_Z^2}$, as shown
schematically in Fig.\ \ref{fig:h+tau-modes}(a); otherwise, $\eps_-(k)$ rises
monotonically from $\eps_-(0) = \Ezero -\EZ$, with a small-$k$ behavior that is
quartic in $k$ for $\EZ = 2 \ER$ but quadratic for any $\EZ > 2 \ER$ [see
Fig.\ \ref{fig:h+tau-modes}(b)].

The helicity- and angular-momentum-resolved densities of states entering
Eq.\ \eqref{constant_V} become
\begin{widetext}
\begin{equation}
\label{rho_h,tau}
\rho_{h,\tau}(\eps) = \begin{cases}
  \rhoz \dfrac{\ER - \tau\EZ}{\sqrt{\ER(\eps-\Emin)}} \:
  \Theta(2\,\ER -\EZ) \: \delta_{h,-}
    & \text{for } \Emin <\eps < \Ezero -\EZ, \\[3ex]
  \dfrac{\rhoz}{2} \biggl[ 1
    - h \, \dfrac{\ER-\tau\EZ}{\sqrt{\ER(\eps-\Emin)}} \biggr]
    & \text{for } \Ezero + h\EZ < \eps < D_h. \\[3ex]
  0 & \text{otherwise,}
\end{cases}
\end{equation}
\end{widetext}
where $D_h = D + 2 h \sqrt{\ER(D-\ER-\Emin)}$.
For weak Zeeman splittings $\EZ \le 2\,\ER$, $\rho_{-,\tau}(\eps)$ features a
van Hove singularity at $\eps=\Emin$, associated with the minimum in
$\epsilon_-(k)$. By contrast, for strong Zeeman splittings $\EZ > 2 \ER$,
there is no divergence of any $\rho_{h,\tau}(\eps)$.

Figures \ref{fig:h+tau-modes} also shows schematic plots of the
angular-momentum-resolved densities of states $\rho_{\tau}(\eps) =
\sum_h \rho_{h,\tau}(\eps)$, which determine $\Gamma_{\tau}(\eps)$ entering
Eq.\ \eqref{H_1chan:Z}. These densities of states coincide with $\rho_0(\eps)$
over the energy range $\Ezero+\EZ < \eps < D_-$. For $\EZ \le 2 \ER$, each
density of states, plotted schematically in Fig.\ \ref{fig:h+tau-modes}(c),
inherits a van Hove singularity at $\eps=\Emin$ from the divergence of
$\rho_{-,\tau}(\eps)$. In addition, $\rho_{-1/2}(\eps)$ exhibits a jump as
the energy drops below $\eps=\Ezero-\EZ$ due to the onset of contributions
from $h=-$ states close to $k=0$. [There is no corresponding jump in
$\rho_{1/2}(\eps)$ because $\beta_{-1}(k)$ vanishes at $k=0$.]
For $\EZ > 2 \ER$, by contrast, $\rho_{1/2}(\eps)$ approaches zero and
$\rho_{-1/2}(\eps)$ rises smoothly to a constant limiting value as
$\eps$ approaches $\Ezero -\EZ$ from above, and both densities of states vanish
for $\eps<\Ezero -\EZ$, as sketched in Fig.\ \ref{fig:h+tau-modes}(d).

\section{Numerical Results}
\label{sec:results}

In order to study Hamiltonians \eqref{H_2chan}, \eqref{H_1chan}, and
\eqref{H_1chan:Z} with the densities of states defined in Eqs.\ \eqref{rho_h}
and \eqref{rho_h,tau}, we have applied the numerical renormalization-group (NRG)
method for the solution of the Anderson model \cite{Krishna-murthy:1980}, as
adapted to treat arbitrary densities of states \cite{Bulla:1997,GBI:1998}. We
set the Wilson discretization parameter to $\Lambda=2.5$, retaining at least
2\,000 many-body states after each iteration. Results are shown for
$\Ezero=-0.08D$ and various combinations of $U$, $\Ed$, and the hybridization
width $\Gamma\equiv\pi\rhoz V^2$. Any value of $\Gamma$ employed in an NRG
calculation should be equivalent in the continuum limit to a hybridization
width $\Gamma_{\eff}=\Gamma/A_{\Lambda}$, where
$A_{\Lambda} = \half (\ln\Lambda) (\Lambda + 1)/(\Lambda - 1)\simeq 1.069$
accounts for a reduction in the density of states that arises from the NRG
discretization \cite{Krishna-murthy:1980}.

The results shown in Sec.\ \ref{subsubsec:spin-correlations} were
obtained by solving the two-channel Anderson model [Eq.\ \eqref{H_2chan}].
All other data presented in this section come from calculations performed on a
one-channel model [either Eq.\ \eqref{H_1chan} or Eq.\ \eqref{H_1chan:Z}].

\subsection{Results without a Zeeman field}
\label{subsec:nofield}

\subsubsection{Thermodynamic properties}
\label{subsubsec:thermo}

\begin{figure}[t]
\centerline{\includegraphics[width=\columnwidth]{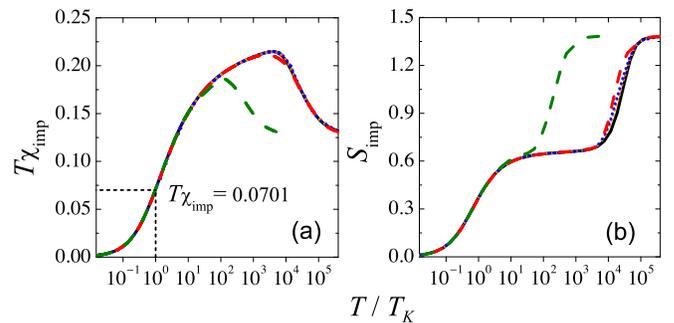}}
\caption{\label{fig:thermo} (Color online)
Impurity contribution to (a) temperature times magnetic susceptibility
$T\chi_{\imp}$, and (b) entropy $S_{\imp}$, both plotted vs scaled temperature
$T/T_K$ for five different cases: a no-Rashba reference (black solid line), and
Rashba energies $\ER/D=0.04$ (red and orange curves) and $0.08$ (green
and blue curves) under the scenarios of fixed Fermi energy (dotted lines)
and constant band filling (dashed lines). The two dotted lines lie directly
on top of one another on the scale of this plot.
The Kondo temperature $T_K$ is as defined in Eq.\ \eqref{T_K:def}.
Data are for $\Ezero=-0.08D$, $U=-2\Ed=0.1D$, and $\Gamma=0.005D$.}
\end{figure}

We begin by showing the temperature variation of the impurity contribution to
two thermodynamic quantities calculated using the effective one-channel
Anderson model [Eq.\ \eqref{H_1chan}]: the magnetic susceptibility
$\chi_{\imp}$ (calculated for equal impurity and band $g$ factors, $g_i=g_b$)
and the entropy $S_{\imp}$. To reduce NRG discretization errors, we employed
interleaved averaging \cite{Oliveira-Oliveira:1994} over three different band
discretizations.

In order to investigate the universality of the low-temperature physics,
Fig.\ \ref{fig:thermo} shows $T\chi_{\imp}$ and $S_{\imp}$ for a
symmetric impurity ($U=-2\Ed=0.1D$, $\Gamma=0.005D$) as functions of $T/T_K$.
Note that we have defined $\chi_{\imp}$ to be the impurity contribution to
the static part of the correlation function for the $z$ component of the total
angular momentum, which reduces to the customary static spin susceptibility in
the case $\epsilon_R=0$. The Kondo temperature $T_K$ was determined via the
conventional criterion \cite{units,Krishna-murthy:1980}
\begin{equation}
\label{T_K:def}
T_K \, \chi_{\imp}(T_K) = 0.0701 ,
\end{equation}
or equivalently, via the condition $S_{\rm imp} (T_K) = 0.383$.
The fact that for $T\lesssim T_K$ the curves for $\ER/D=0.04$ and $0.08$,
calculated both for fixed Fermi energy (dotted lines) and for constant
band-filling (dashed lines), lie on top of the curve for $\ER=0$ (solid line)
provides evidence that the low-temperature thermodynamic properties are those
of a conventional Kondo effect, with $\lim_{T\to 0}T\chi_{\imp}=0$ and
$\lim_{T\to 0}S_{\imp}=0$ indicating complete ground-state screening of the
impurity degree of freedom.

Even well above the Kondo temperature, four of the five curves in each panel
of Fig.\ \ref{fig:thermo} are virtually indistinguishable, exhibiting both a
high-temperature free-impurity regime (in which $T\chi_{\imp}\simeq 1/8$,
$S_{\imp}\simeq \ln 4$) and an intermediate-temperature local-moment regime
(where $T\chi_{\imp}\simeq 1/4$, $S_{\imp}\simeq \ln 2$). The only exception
is the curve for $\ER=0.08D$ at constant band filling. Here, the Fermi energy
is pushed down into the van Hove singularity in the effective one-channel
density of states. The enhanced hybridization width drives the system from
its local-moment regime into mixed valence, where $T\chi_{\imp}$ never rises
close to $1/4$ and with decreasing temperature $S_{\imp}$ drops from $\ln 4$
to $0$ with at most a weak shoulder around $\ln 2$.

\subsubsection{Kondo temperature}
\label{subsubsec:TK}

\begin{figure}[t]
\centerline{\includegraphics[width=\columnwidth]{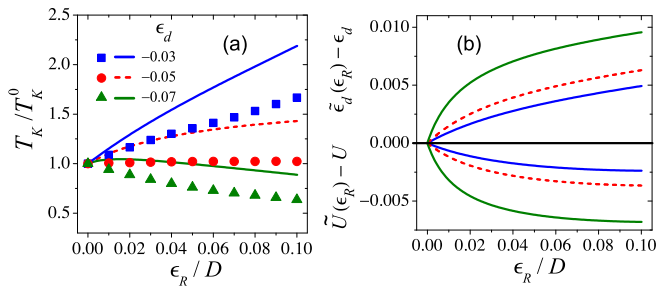}}
\caption{\label{fig:T_K:const-E_F} (Color online)
(a) Scaled Kondo temperature $T_K/T_K^0$ vs Rashba energy $\ER$ for fixed Fermi
energy $\EF=0$ and for different values of $\Ed$ expressed in the legend in
units of $D$. Symbols represent NRG data and lines are the result of the
perturbative treatment described in the text.
(b) Perturbative shifts $\tEd-\Ed$ and $\tU-U$ in the effective impurity
parameters vs $\ER$ for the same cases shown in (a). 
Data are for $\Ezero=-0.08D$, $U=0.1D$, and $\Gamma=0.005D$.}
\end{figure}	

\begin{figure}[tb]
\centerline{\includegraphics[width=\columnwidth]{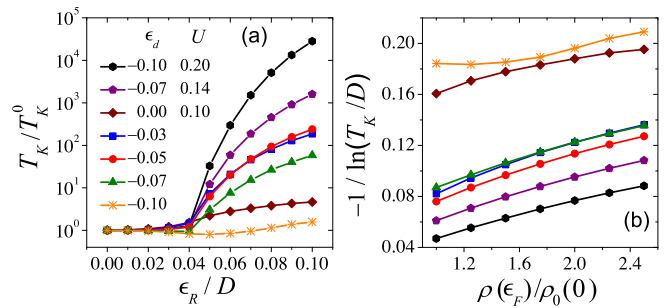}}
\caption{\label{fig:T_K:const-eta} (Color online)
Kondo temperature for a constant band filling fraction $\eta\simeq 0.074$:
(a) $T_K/T_K^0$ vs $\ER$ for different values of $\Ed$ and $U$ expressed in
the legend in units of $D$;
(b) results for $\ER\ge 0.04D$ plotted as $-1/\ln(T_K/D)$ vs
$\rho[\EF(\ER)]/\rho_0(0)$.
Data are for $\Ezero=-0.08D$ and $\Gamma=0.005D$.}
\end{figure}

Given that the low-temperature physics is characterized by a single
scale $T_K$, we now examine more closely the effect of Rashba coupling on the
Kondo temperature. Figures \ref{fig:T_K:const-E_F}(a) and
\ref{fig:T_K:const-eta}(a) plot the ratio of the Kondo temperature $T_K$ at
Rashba energy $\ER$ to its value $T_K^0$ in the absence of SO interaction.

Figure \ref{fig:T_K:const-E_F} treats the scenario of fixed Fermi energy
$\EF = 0$ for $\Ezero=-0.08D$, $U=0.1D$, and $\Gamma=0.005D$.
The Kondo temperature in the absence of SO interaction varies
with $\Ed$, taking its smallest value around $\Ed = -U/2 = -0.05D$, for
which case $T_K^0 \simeq 1.7\times 10^{-6}D$. This is very close to the
value $1.4\times 10^{-6}D$ given by substituting
$\Gamma(\EF)=\Gamma_{\eff}$ into Haldane's estimate
\cite{Haldane:1978a,units}
\begin{equation}
\label{Haldane}
T_K \simeq 0.29 \sqrt{U \, \Gamma(\EF)}
  \exp\biggl[\frac{\pi \Ed (U+\Ed)}{2U\,\Gamma(\EF)} \biggr]. 
\end{equation}
for the Kondo temperature of an Anderson impurity in the limit that
$0 < \Gamma(\EF) \ll U+\Ed, -\Ed\ll D$.

The data symbols in Fig.\ \ref{fig:T_K:const-E_F}(a) show $T_K$ to be only
weakly affected by Rashba coupling, just as was found in Ref.\
\onlinecite{Zitko-Bonca:2011}. The Kondo scale displays a quasi-linear
$\ER$ dependence \cite{Ivanov:2012} with a slope that is positive for
$\Ed <-U/2$, negative for $\Ed>-U/2$, and essentially vanishing for
$\Ed=-U/2$ (also in agreement with Ref.\ \onlinecite{Zitko-Bonca:2011}). Since
the effective density of states $\rho(\eps)$ is independent of $\ER$ in a
window of energies around $\EF$, any modification of $T_K$ must arise from
changes in the total density of states near the band edges. 

One can attempt to analyze the effect of density of states changes via
a perturbative treatment along the lines of Haldane's derivation of poor-man's
scaling equations for the Anderson impurity model \cite{Haldane:1978}. This
treatment, described further in Appendix \ref{app:pert}, can be used to map
the Anderson impurity model with density of states $\rho(\eps)$ onto another
Anderson impurity model with the no-Rashba density of states $\rho_0(\eps)$,
but with the impurity parameters $\Ed$ and $U$ replaced by modified values
$\tEd$ and $\tU$ chosen so as to preserve (approximately) the same low-energy
impurity properties as the original model. As explained in Appendix
\ref{app:pert}, the hybridization matrix element $V$ remains unchanged under
this approach.

For purely quadratic band dispersion in the absence of Rashba SO interaction,
expressions for the renormalized parameter $\tEd$ and $\tU$ can be obtained
in closed form, but they are too cumbersome to reproduce here. Instead, Fig.\
\ref{fig:T_K:const-E_F}(b) shows the evolution with $\ER$ of the shifts
$\tEd-\Ed$ and $\tU-U$ for each of the three $\Ed$ values illustrated in Fig.\
\ref{fig:T_K:const-E_F}(a). An upward shift in the level energy and a downward
shift in the on-site interaction both grow with $\ER$ and with $-\Ed$, becoming
10\% corrections in the most extreme case shown. Upon substitution into the
Haldane formula [Eq.\ \eqref{Haldane}], these shifts in $\tEd$ and $\tU$ have
opposite effects on $T_K$, so any overall change in $T_K$ is the result of a
subtle balance. The predicted curves for $T_K/T_K^0$ vs $\ER$ [solid lines in
Fig.\ \ref{fig:T_K:const-E_F}(a)] display the correct trends with growing
$-\Ed$, but the fact that NRG and perturbative curves for the same value of
$\Ed$ are not in close correspondence is an indication of the delicacy of the
interplay between the parameter renormalizations.

Figure \ref{fig:T_K:const-eta}(a) shows very different behavior in cases where
the band filling is fixed. For weak SO interaction, the total density of states
$\rho(\eps)$ near the Fermi energy remains independent of $\ER$, and $T_K$ has
a quasi-linear behavior similar to that found for fixed $\EF$. However, once
$\ER>|\Ezero|/2$, the system enters the helical metal regime and both
$\rho(\EF)$ and $\Gamma(\EF)$ rise rapidly. This rise is magnified in the Kondo
temperature due to the exponential dependence of $T_K$ on $\Gamma(\EF)$ shown
in Eq.\ \eqref{Haldane}. Figure \ref{fig:T_K:const-eta}(b) confirms such a
dependence of the numerically determined value of $T_K$. The five lower curves
exemplify the Kondo regime, where the weak deviations from linearity in this
plot of $-1/\ln(T_K/D)$ vs $\rho(\EF)$ can be attributed primarily to the
$\sqrt{\rho(\EF)}$ prefactor in Eq.\ \eqref{Haldane}. The remaining two cases
($U=0.1D$, $\Ed=0$ and $\Ed=-0.1D$) correspond to mixed valence, where the
low-temperature scale is no longer expected to depend exponentially on
$\Gamma(\EF)$.

\subsubsection{Static angular-momentum correlations}
\label{subsubsec:spin-correlations}

\begin{figure}[tb]
\centerline{\includegraphics[width=0.95\columnwidth]{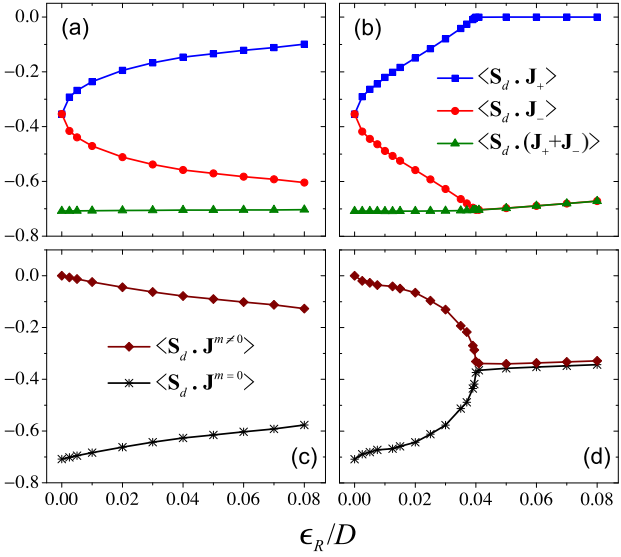}}
\caption{\label{fig:corr} (Color online)
Static correlations between the impurity spin and the total angular momentum in
different conduction-band channels, calculated for (a,c) fixed Fermi energy
$\ER=0$, or (b,d) constant band filling $\eta\approx 0.074$.
The upper panels show $\langle\Sd\cdot\J_h\rangle$ for helicity $h=\pm$, while
the lower panels plot the correlation of $\Sd$ with $\J^{m=0}$ (the total
angular momentum of all electrons with orbital angular momentum $m=0$) and with
$\J^{m\ne 0}$ defined in Eq.\ \eqref{J^m!=0:def}.
Data are for $\Ezero=-0.08D$, $U=-2\Ed=0.1D$, and $\Gamma=0.005D$.}
\end{figure}	

Further insight can be gained into the nature of Kondo physics in presence of
Rashba SO interaction by studying the correlations between the impurity spin
\begin{equation}
\label{S_d:def}
\Sd = \half\sum_{\tau,\tau'} d_{\tau}^{\dag} \: \bm{\sigma}_{\tau,\tau'} \:
      d_{\tau'}^{\pdag}
\end{equation}
and the total angular momentum
\begin{equation}
\label{J_h:def}
\J_h
  = \half\sum_{\tau,\tau'} \int\!dk \: \tc_{k,h,\tau}^{\dag} \:
    \bm{\sigma}_{\tau,\tau'} \: \tc_{k,h',\tau'}^{\pdag}
\end{equation}
of all conduction-band electrons that have helicity $h$ and angular momentum
$z$ component $\pm\half$. The appendix provides details of how such
correlations can be obtained within the NRG treatment of the two-channel
Anderson Hamiltonian [Eq.\ \ref{H_2chan}]. The results presented below are
static values calculated in the limit of absolute temperatures $T\to 0$.

Figure \ref{fig:corr}(a) plots $\langle \Sd\cdot \J_h\rangle$ vs $\ER$ under
the scenario where the Fermi energy is fixed at $\EF=0$. At zero Rashba energy,
$\langle \Sd\cdot \J_h\rangle$ is the same for helicities $h=\pm$. As $\ER$
increases, the impurity spin becomes more strongly correlated with the $h=-$
channel and less strongly with the $h=+$ electrons. This can be understood as
a density of states effect because the dimensionless ratio
\begin{equation}
\label{r_h}
r_h=\langle \Sd\cdot \J_h\rangle/[D\rho_h(\EF)]
\end{equation}
(not plotted) turns out to be almost independent of $\ER$ and $h$. Since the
combined Fermi-level density of states $\rho(\EF)=\rho_+(\EF)+\rho_-(\EF)$
remains constant, it is therefore unsurprising that
$\langle\Sd\cdot(\J_+ + \J_-)\rangle$ barely changes with $\ER$.

In cases where the band filling is constant [e.g., Fig.\ \ref{fig:corr}(b)], the
behavior found for Rashba energies $\ER<-\Ezero/2$ is qualitatively the same as
for fixed $\EF$. However, once the helical regime is reached ($\ER>-\Ezero/2$),
$\rho_+(\EF)=0$ and $\langle \Sd\cdot \J_{+}\rangle$ almost vanishes; the
impurity spin is almost exclusively correlated with the $h=-$ channel. In this
parameter range, the growth of $\rho_-(\EF)$ inside the helical regime increases
the occupancy of the empty and doubly occupied impurity states and decreases the
local-moment character, leading to a gradual decline in the magnitude of
$\langle\Sd\cdot(\J_+ + \J_-)\rangle$ with increasing $\ER$. The
contrast with the fixed-$\EF$ scenario is highlighted by the facts that
$r_+=\infty$ due to the numerator on the right-hand side of Eq.\ \eqref{r_h}
being small but nonzero due to correlation of $\Sd$ with high-energy electrons,
while $r_-$ rapidly approaches zero with increasing $\ER$ due to
$\langle\Sd\cdot \J_{-}\rangle$ being nearly saturated but $\rho_-(\EF)$ still
rising.

Figures \ref{fig:corr}(c) and \ref{fig:corr}(d) separate (for the fixed-$\EF$
and constant-$\eta$ scenarios, respectively) two contributions to the overall
impurity-band angular momentum correlation $\langle\Sd\cdot(\J_+ +\J_-)\rangle$:
(i) $\langle\Sd\cdot\J^{m=0}\rangle$, where $\J^{m=0}$ is the
total angular momentum of all electrons with orbital angular momentum $m=0$,
and (ii) $\langle\Sd\cdot\J^{m\ne 0}\rangle$, where
\begin{equation}
\label{J^m!=0:def}
\J^{m\ne 0} = \J_+ + \J_- -\J^{m=0} .
\end{equation}
In both panels, the most striking result is the appearance for $\ER>0$ of a
nonzero correlation $\langle\Sd\cdot\J^{m\not=0}\rangle$ that arises, not
through direct interaction with the impurity (which is confined to $m=0$), but
rather indirectly through Rashba mixing of $m=0$ states with $m=\pm 1$ states.

The details of Figs.\ \ref{fig:corr}(c) and \ref{fig:corr}(d) can be understood
in terms of the helicity-resolved densities of states. Equation \eqref{a_kht}
shows that the operators $\tc_{k,h,\tau}$ entering Eq.\ \eqref{H_2chan:k-space}
contain $m=0$ and $m=\pm 1$ components of equal magnitude with a relative phase
that is opposite for $h=+$ and $h=-$. For $\ER=0$, the two helicity channels
participate equally in Kondo screening in a manner that produces complete
destructive interference of correlations contributing to
$\langle\Sd\cdot\J^{m\not=0}\rangle$. With increasing $\ER$ there is a
growing difference $\rho_-(\eps)-\rho_+(\eps)$ for energies
$\eps$ near $\EF$, leading to imperfect cancellation of correlations
between the impurity and $m=\pm 1$ electrons and a gradual convergence of
$\langle\Sd\cdot\J^{m=0}\rangle$ and $\langle\Sd\cdot\J^{m\not=0}\rangle$. Once
the system enters the helical regime [$\ER>-\Ezero/2=0.04D$ in Fig.\
\ref{fig:corr}(d)], $\rho_+(\EF)=0$, and the two correlation measures differ
only due to contributions from electrons far from the Fermi energy.

\subsection{Results with a Zeeman field}
\label{subsec:Zfield}

In order to study the combined effect of Rashba and Zeeman couplings on Kondo
correlations, we have calculated the zero-temperature impurity
polarization $\Mimp = \langle J_z\rangle - \langle J_z\rangle_0$, where $J_z$
is the $z$ component of the total system angular momentum, and
$\langle\ldots\rangle$ and $\langle\ldots\rangle_0$ denote, respectively,
expectation values in the presence and absence of the impurity. $\Mimp$ is
the natural extension of the usual impurity spin magnetization
$\langle S_z\rangle - \langle S_z\rangle_0$, to which it reduces
in the absence of Rashba coupling (i.e., for $\ER=0$).

\begin{figure}[tb]
\centerline{\includegraphics[width=\columnwidth]{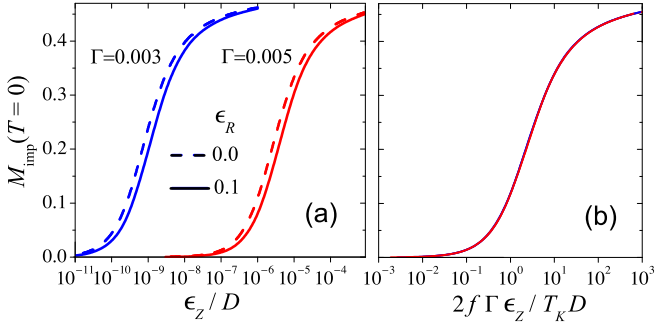}}
\caption{\label{fig:MvsEz:const-E_F} (Color online)
(a) Zero-temperature impurity polarization $\Mimp$ vs Zeeman energy $\EZ$ for
$\Ezero=-0.08D$, $U=-2\,\Ed=0.15D$, and for the values of $\Gamma$ and $\ER$
labeled on the plot in units of $D$.
(b) Same data as in (a), replotted vs $2 f \Gamma\EZ/T_K D$, where
$f=2.4$ and $1.8$ for $\ER/D=0$ and $0.1$, respectively.}
\end{figure}

Initially, we focus on the scenario of fixed Fermi energy $\EF = 0$. Figure
\ref{fig:MvsEz:const-E_F}(a) shows the zero-temperature impurity polarization
$\Mimp$ as a function of $\EZ$ for $U=-2\,\Ed=0.15D$, for $\ER/D=0$ and $0.1$, 
and for $\Gamma/D=0.003$ and $0.005$. The impurity parameters are such that for
$\ER=0$, the Kondo temperatures
$T_K^0(\Gamma=0.003D)\approx 4.4\times 10^{-12}D$ and
$T_K^0(\Gamma=0.005D)\approx 2.6\times 10^{-8}D$ differ by four orders of
magnitude. As expected, $\Mimp$ is in all cases an increasing function of
$\EZ$. For a given value of $\EZ>0$, $\Mimp$ is a decreasing function of
$\ER$. This can be understood by noting that the Zeeman field enters Eq.\
\eqref{rho_h,tau} in the combination $\ER-\tau\EZ$.  Therefore, the splitting
of the density of states for $\tau=\pm\half$ becomes less significant with
increasing Rashba coupling.

Figure \ref{fig:MvsEz:const-E_F}(b) demonstrates a very good collapse of the
polarization data in Fig.\ \ref{fig:MvsEz:const-E_F}(a) when plotted as a
function of $2 f \Gamma\EZ / T_K D$, where $T_K$ is the Kondo temperature for
$\EZ=0$ and $f$ is a dimensionless fitting parameter that depends on $\ER$. It
is particularly notable that for given $\ER$ (including the case of no Rashba
coupling), the Zeeman field is scaled by $T_K/\Gamma$ rather than by the Kondo
temperature itself, as would be the case for a nonzero impurity $g$ factor.
Small deviations from the scaling collapse occur only in the parameter range
$\EZ\gtrsim\Gamma$.

The scaling shown in Fig.\ \ref{fig:MvsEz:const-E_F}(b) can be understood via
a second application of the perturbative treatment outlined in Appendix
\ref{app:pert}. One can regard the Zeeman field as introducing a spin-dependent
change $\Delta\rho_{\tau}=\rho_{\tau}(\eps)-\rho(\eps)$ in the density of
states of electrons with angular momentum $z$ component $\tau$, where
$\rho(\eps)=\rho_+(\eps)+\rho_-(\eps)$ is derived from Eq.\ \eqref{rho_h}. The
perturbative method maps the problem to a one-channel Hamiltonian of the form
of Eq.\ \eqref{H_1chan} in which both spin species share the same density of
states $\rho(\eps)$ and to lowest order in $\EZ$ and $\Gamma$, the on-site
Coulomb repulsion $U$ and the hybridization width $\Gamma$ remain unchanged,
but the impurity level energy becomes $\tau$ dependent \cite{FM-lead-splitting}:
\begin{equation}
\label{tildeE_tau}
\tilde{\eps}_{d,\tau} \simeq \Ed - 2 \tau f(\ER,U,\Ed)\, \Gamma \EZ/D .
\end{equation}
In other words, the Zeeman coupling effectively spin-splits the impurity level
by an amount proportional to $\Gamma\EZ$. Just as would be the case if this
splitting were caused by a magnetic field that coupled directly to the impurity
spin, the Kondo correlations are destroyed when the splitting becomes
comparable to $T_K$. In the present context, the result is a scaling dependence
on the dimensionless quantity $\Gamma\EZ/D T_K$ rather than the conventional
(i.e., $g_i \ne 0$) combination $\EZ/T_K$. This conclusion is not restricted
to situations involving Rashba SO interaction, and holds quite generally for
realizations of the Anderson impurity model with $g_i=0$.

Equation \eqref{tEds:def} yields an algebraic expression for $f$ to lowest
order in $\EZ/\ER$ that is rather cumbersome and will not be reproduced here.
More generally, one can consider a combination of a bulk Zeeman splitting
and a local field $B_{\mathrm{loc}}$ that couples only to the impurity. After
integrating out the Zeeman field, one arrives at a renormalized impurity
energy\ $\tilde{\eps}_{d,\tau} = \Ed - 2 \tau f(\ER,U,\Ed)\, \Gamma \EZ/D
+ \tau g_i B_{\mathrm{loc}}$. This allows the numerical determination of $f$ as
$f=g_i B_{\mathrm{comp}}/(2\Gamma\EZ)$, where $B_{\mathrm{comp}}$ is the
compensation value of the local field such that
$\tilde{\eps}_{d,\tau} = \Ed$ and hence $\Mimp=0$.

\begin{figure}[tb]
\centerline{\includegraphics[width=\columnwidth]{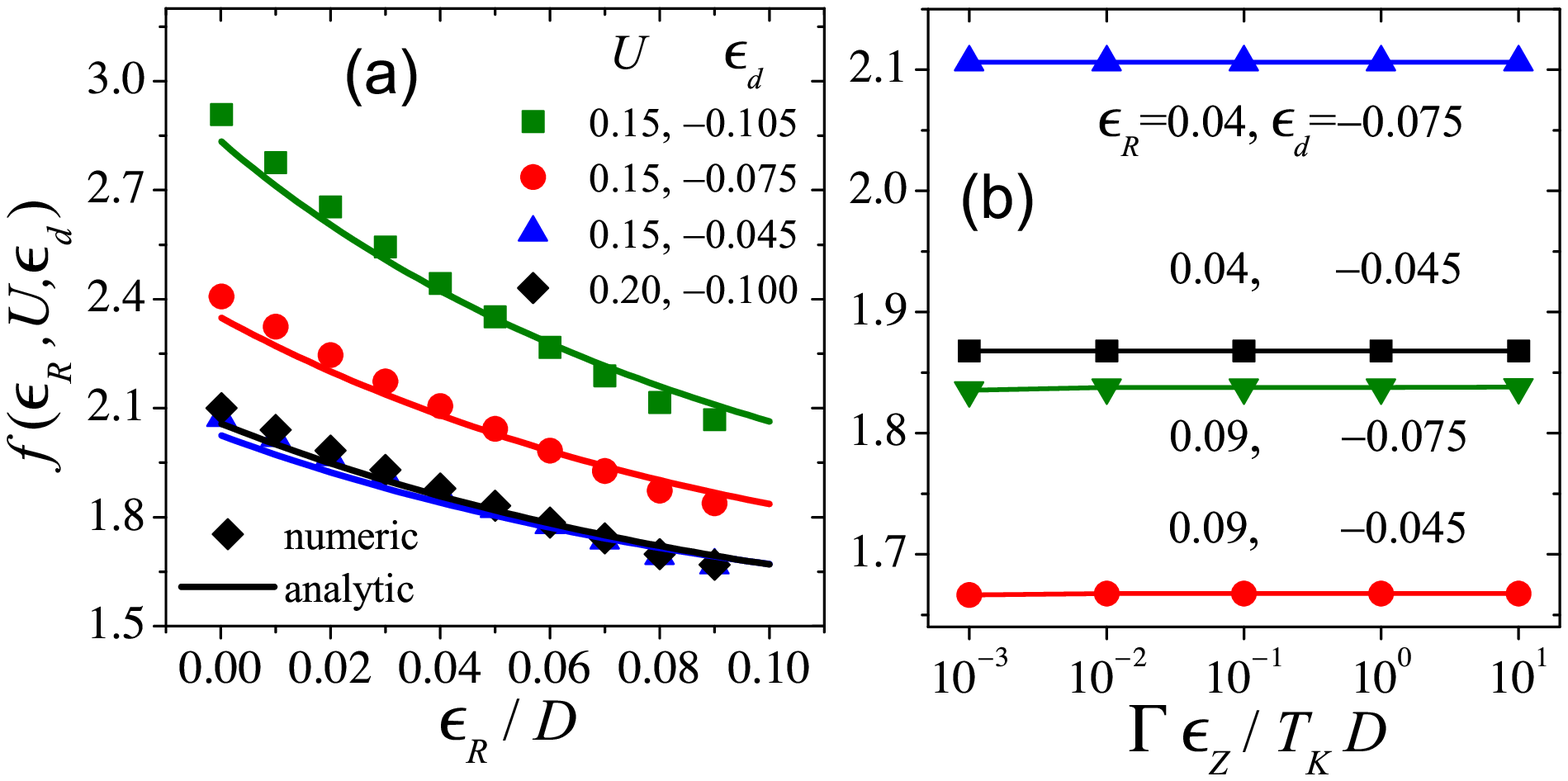}}
\caption{\label{fig:f:const-E_F} (Color online).
Quantity $f$ entering the scaling collapse of the impurity
polarization for a system with fixed Fermi energy $\EF=0$:
(a) $f$ vs $\ER$ for different values of $U$ and $\Ed$ expressed in the legend
in units of $D$. Each symbol was obtained via an NRG calculation of the
compensating local field for $\Gamma\EZ/T_K D=0.1$, while the lines represent
algebraic results based on Eq.\ \eqref{tildeE_tau}.
(b) NRG values of $f$ vs $\Gamma \EZ/T_K D$ for $U=0.15D$ and different values
of $\ER$ and $\Ed$ expressed in the legend in units of $D$. Lines are guides to
the eye showing that $f$ is independent of $\EZ$ for
$\Gamma \EZ/ T_K D\lesssim 10$.
Data in both panels are for $\Ezero=-0.08D$ and $\Gamma=0.005D$.}
\end{figure}

Figure \ref{fig:f:const-E_F}(a) illustrates the behavior of $f$ vs $\ER$ for
different combinations of $U$ and $\Ed$. All cases show a monotonic decrease of
$f$ with increasing $\ER$. There is also an interplay between $U$ and $\Ed$,
such that the $f$ values for $(U/D,\Ed/D)=(0.2,-0.1)$ and $(0.15,-0.045)$ lie
almost on top of each other. In all the cases shown, there is good quantitative
agreement between the numerical and algebraic values plotted using symbols and
lines, respectively. 

Figure \ref{fig:f:const-E_F}(b) shows $f$ as a function of $\Gamma \EZ/ T_K D$
for $U=0.15D$ and a few selected values of $\ER$ and $\Ed$. The most salient
feature is that $f$ remains constant as $\Gamma \EZ/T_K D$ is varied over
four orders of magnitude, indicating the lack of dependence on $\Gamma$
and/or $\EZ$. Deviations from universality are again seen only in the regime
of very large Zeeman fields where the impurity polarization approaches
saturation at $\Mimp=0.5$.

\begin{figure}[tb]
\centerline{\includegraphics[width=\columnwidth]{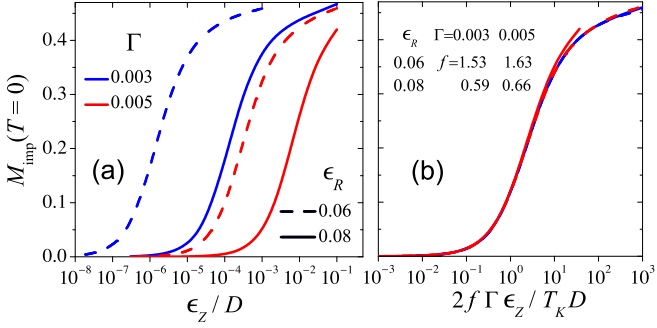}}
\caption{\label{fig:MvsEz:const-eta} (Color online)
(a) Zero-temperature impurity polarization $\Mimp$ vs Zeeman energy $\EZ$ for
$\Ezero=-0.08D$, $U=-2\,\Ed=0.15D$, and for the values of $\Gamma$ and $\ER$
labeled on the plot in units of $D$.
(b) Same data as in (a), replotted vs $2 f \Gamma\EZ/T_K D$ with values of
$f$ shown in the legend. In this case, $f$ depends on $\Gamma$ as well as
$\ER$.}
\end{figure}

\begin{figure}[tb]
\centerline{\includegraphics[width=\columnwidth]{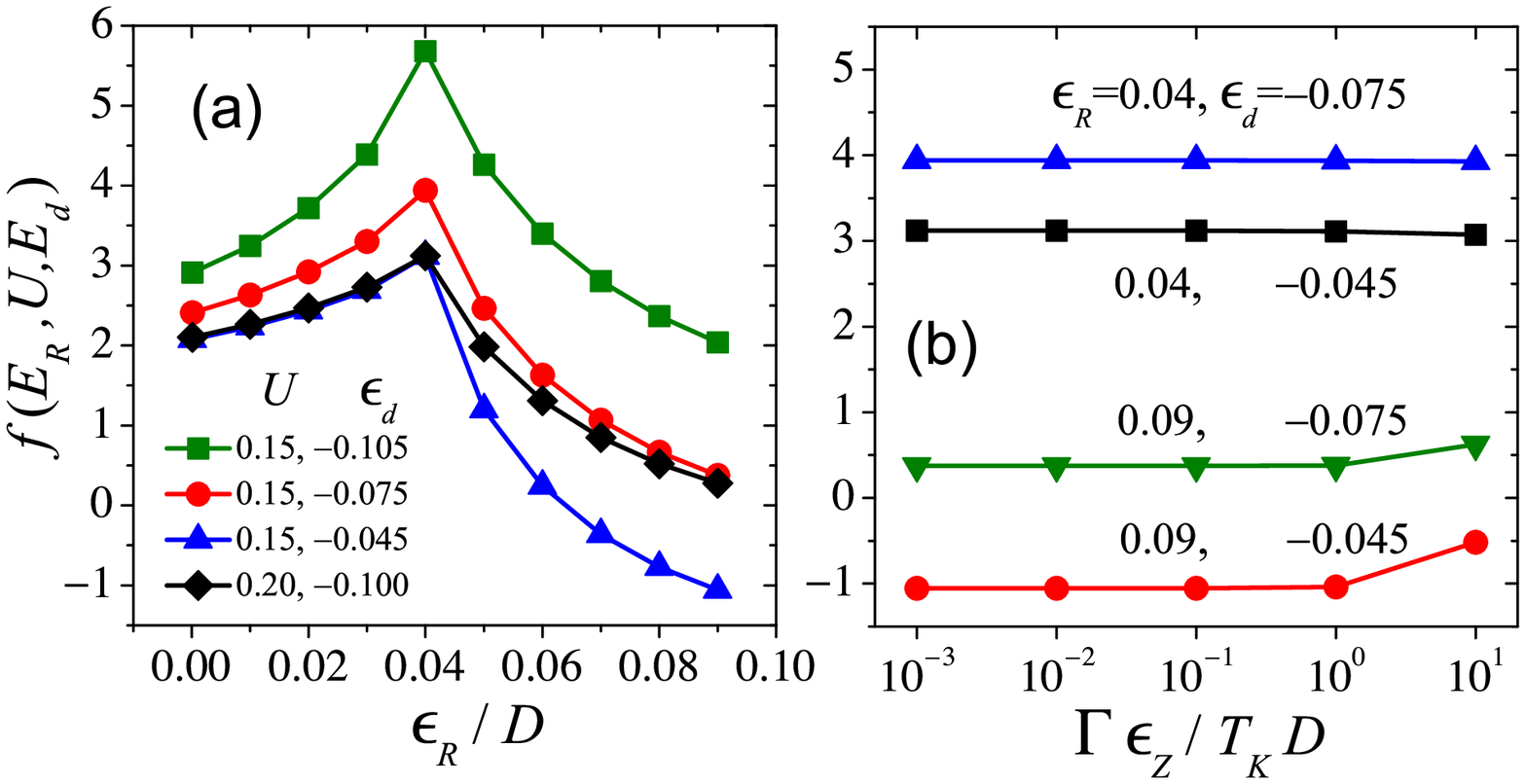}}
\caption{\label{fig:f:const-eta} (Color online)
Quantity $f$ entering the scaling collapse of the impurity
polarization for a system with constant band filling $\eta\simeq 0.074$:
(a) $f$ vs $\ER$ for different values of $U$ and $\Ed$ expressed in the legend
in units of $D$. Each symbol was obtained via an NRG calculation of the
compensating local field for $\Gamma\EZ/T_K D=0.1$, while the lines are guides
to the eye.
(b) NRG values of $f$ vs $\Gamma \EZ/T_K D$ for $U=0.15 D$ and different values
of $\ER$ and $\Ed$ expressed in the legend in units of $D$. Line are guides to
the eye showing that $f$ is independent of $\EZ$ for
$\Gamma \EZ/ T_K D\lesssim 10$.
Data in both panels are for $\Ezero=-0.08D$ and $\Gamma=0.005D$.}
\end{figure}

Finally, we consider the effect of Zeeman splitting under the scenario of
constant band filling. Figure \ref{fig:MvsEz:const-eta}(a) shows $\Mimp$ vs
$\EZ$ for $U=-2\,\Ed=0.15D$, for $\ER/D=0.06$ and $0.08$, and for
$\Gamma/D=0.003$ and $0.005$. Since Rashba SO interaction changes
the Fermi-energy density of states, the Zeeman energy needed to destroy Kondo
correlations is exponentially sensitive to both $\ER$ and $\Gamma$. As under
the fixed-$\EF$ scenario, all the polarization curves share a similar shape
(except for $\EZ \gtrsim\Gamma$) and a universal scaling dependence
on $2 f\Gamma\EZ/T_K D$ is confirmed in Fig.\ \ref{fig:MvsEz:const-eta}(b).
In this case, however, the parameter $f$ depends not just on $\ER$, $U$, and
$\Ed$, but also decreases with $\Gamma$. Under the constant-filling scenario,
the Fermi level lies in an energy range where the density of states is
spin-split, so a perturbative approach to first order in $\Gamma$ is likely
insufficient to reproduce the parameter dependences of $f$. For this reason, we
focus on the numerical estimation of $f$ via the compensation field. Figure
\ref{fig:f:const-eta} plots the results for the same parameters as were used
in Fig.\ \ref{fig:f:const-E_F}. Note that $f$ exhibits a nonmonotonic
dependence on $\ER$, with a maximum at $\ER=-\Ezero/2=0.04D$. For
$\ER\le -\Ezero/2$, the value of $f$ under the constant-filling scenario is
generally greater than for fixed Fermi energy. For $\ER> -\Ezero/2$, the
range where the system is in its helical regime for $\EZ=0$, $f$ decreases
rapidly with increasing Rashba coupling and may even become negative.

\section{Summary}
\label{sec:summary}

We have studied the effect of bulk Rashba SO interaction on Kondo correlations
between a magnetic impurity and a two-dimensional electron gas. The
low-temperature thermodynamic properties exhibit the conventional Kondo form
with scaling in terms of $T/T_K$, providing evidence for complete quenching of
the impurity degree of freedom as $T\to 0$.
In most situations, the Kondo temperature $T_K$ is little affected by the Rashba
coupling, as has been pointed out previously \cite{Zitko-Bonca:2011}. However,
within a helical regime that can in principle be accessed for high Rashba
couplings and/or low electron densities, the Kondo temperature exhibits an
exponential enhancement compared to the situation without Rashba interaction. 

Our analysis of static angular-momentum correlations demonstrates and
quantifies an indirect, Rashba-induced coupling of the impurity spin with
conduction channels of nonzero orbital angular momentum about the impurity site.
This coupling can be regarded as a manifestation of a Dzyaloshinskii-Moriya term
found previously by mapping the Anderson impurity model with bulk Rashba SO
interaction to an effective Kondo model \cite{Zarea:2012}. A perturbative
renormalization-group analysis of this Kondo model showed an exponential
enhancement of $T_K$. We should note, however, that the RG equations were
solved neglecting the helicity dependence of the Fermi wave vector. A complete
analysis of the effective Kondo model will be presented elsewhere.

Optical irradiation experiments seem to be good candidates to explore the
helical regime of a Rashba-coupled 2DEG.  Motivated by the proposal of
Ref.\ \onlinecite{Ojanen-Kitagawa:2012}, we have also investigated Kondo
physics in the presence of both Rashba SO interaction and Zeeman splitting
of the bulk electrons (but not of the impurity). The characteristic Zeeman
energy scale for the destruction of the Kondo effect is not $T_K$, as would
be expected in the case of a true magnetic field that couples directly to
the impurity spin, but rather $T_K/\Gamma$ where $\Gamma$ is the impurity
hybridization width. The behavior can be accounted for to reasonable
quantitative accuracy by a perturbative treatment of the Zeeman splitting
of the host density of states.

\begin{acknowledgments}
We acknowledge valuable conversations with M.\ Zarea.
This work was supported in part under NSF Grants No.\ DMR-1107814 and
DMR-1508122 (Florida) and DMR-1108285 and DMR-1508325 (Ohio).
The work of K.I.\ was performed in part at the Aspen Center for Physics,
which is supported by National Science Foundation grant PHY-1066293.
\end{acknowledgments}

\appendix

\section{Perturbative Analysis}
\label{app:pert}

This appendix considers an Anderson impurity model
\begin{multline}
\label{H_1chan:gen}
H = \sum_{\sigma} \int \! d\eps \; \eps \:
    c_{\eps,\sigma}^{\dag} c_{\eps,\sigma}^{\pdag}
  + \sum_{\sigma} \Eds d_{\sigma}^{\dag} d_{\sigma}^{\pdag}
  + U d_{\up}^{\dag} d_{\up}d_{\dn}^{\dag} d_{\dn} \\
  + \sum_{\sigma} V_{\sigma} \int \! d\eps \: \rho_{\sigma}(\eps)
    \: \bigl( c_{\eps,\sigma}^{\dag} d_{\sigma}^{\pdag} + \text{H.c.} \bigr) ,
\end{multline}
where the conduction-band density of states $\rho_{\sigma}(\eps)$, the impurity
level energy $\Eds$, and the hybridization matrix element $V_{\sigma}$ are all
allowed to depend on $\sigma=\,\uparrow,\,\downarrow$. The goal is to construct
an approximate mapping of Eq.\ \eqref{H_1chan:gen} to a similar Hamiltonian
having a different conduction-band density of states $\trho_{\sigma}(\eps)$,
and with the impurity parameters $\tilde{\eps}_{d,\sigma}$, $\tU$, and
$\tV_{\sigma}$ chosen to ensure that the two models share the low-energy physics
(at least as it pertains to the impurity properties).

Haldane, in his derivation of poor-man's scaling equations for the Anderson
model \cite{Haldane:1978}, used perturbation theory in the hybridization matrix
element to take into account the effect of all conduction-band states in a
narrow window of energies near each band edge. Here, we perform a similar
calculation in order to find energy shifts arising from the density-of-states
difference $\Delta\rho_{\sigma}(\eps) =
\rho_{\sigma}(\eps) - \trho_{\sigma}(\eps)$ at all energies $\eps$. Like
Haldane, our focus is on four many-body states $|0\rangle$,
$|\sigma\rangle = d_{\sigma}^{\dag} |0\rangle$, and
$|2\rangle = d_{\up}^{\dag} d_{\dn}^{\dag} |0\rangle$, formed by combining the
conduction-band ground state [having $N_k$ electrons of energy
$\eps(\bk)<\eps_F$] with one of the possible configurations of the Anderson
impurity level. These many-body states have energies $E_0$,
$E_{\sigma}=E_0+\Ed$, and $E_2=E_{\up}+E_{\dn}-E_0+U$, respectively.

The state $|0\rangle$ can decrease its energy through virtual tunneling of an
electron of spin $\sigma$ from a band state below the Fermi level to the
impurity and then back to the original band state. Integrating the contribution
to such processes arising just from the density of states difference
$\Delta\rho_{\sigma}(\eps)$, then summing over $\sigma$, transforms the state
$|0\rangle$ to one $|\tilde{0}\rangle$ having an energy $\tilde{E}_0$ that,
to second order in $V$, is
\begin{equation}
\label{tE_0}
\tilde{E}_0 = E_0 - \sum_{\sigma} V_{\sigma}^2 \int_{-\infty}^{\EF}
  \frac{\Delta\rho_{\sigma}(\eps) \, d\eps}{-\eps+\Eds} \, .
\end{equation}

Similarly, the state $|2\rangle$ can decrease its energy through virtual
tunneling of a spin-$\sigma$ electron from the impurity to a band state above
the Fermi level and then back to the impurity. The contribution to such
processes arising from $\Delta\rho_{\sigma}(\eps)$, when summed over $\sigma$,
transforms $|2\rangle$ to $|\tilde{2}\rangle$ with energy
\begin{equation}
\label{tE_2}
\tilde{E}_2 = E_2 - \sum_{\sigma} V_{\sigma}^2 \int^{\infty}_{\EF}
  \frac{\Delta\rho_{\sigma}(\eps) \, d\eps}{\eps-U-\Eds} \, .
\end{equation}

Lastly, the state $|\sigma\rangle$ can lower its energy through (i) tunneling of
an electron with spin $\sigma$ from the impurity to a band state above the Fermi
level and then back to the impurity, and (ii) tunneling of an electron with
spin $-\sigma$ from a band state below the Fermi energy to the impurity and then
back to the original band state. The contribution to such processes arising from
$\Delta\rho(\eps)$ transforms $|\sigma\rangle$ to $|\tilde{\sigma}\rangle$ with
energy
\begin{equation}
\label{tE_s}
\tilde{E}_{\sigma} = E_{\sigma} - V_{\sigma}^2 \int^{\infty}_{\EF}
  \frac{\Delta\rho_{\sigma}(\eps) \, d\eps}{\eps-\Eds}
  - V_{-\sigma}^2 \int_{-\infty}^{\EF}
  \frac{\Delta\rho_{-\sigma}(\eps) \, d\eps}{-\eps+U+\eps_{d,-\sigma}} \, .
\end{equation}

Equations \eqref{tE_0}--\eqref{tE_s} can be used to define shifted impurity
parameters
\begin{equation}
\label{tEds:def}
\tilde{\eps}_{d,\sigma} = \tilde{E}_{\sigma} - \tilde{E}_0
\end{equation}
and
\begin{equation}
\label{tU:def}
\tU = \tilde{E}_2+\tilde{E}_0-\tilde{E}_{\up} - \tilde{E}_{\dn}.
\end{equation}
Corrections to the hybridization matrix elements (arising as a consequence of
wave-function renormalization) are found to be of order $V^3$, so at the level
of our approximation,
\begin{equation}
\tV_{\sigma} = V_{\sigma}.
\end{equation}
 
Since $\Delta\rho_{\sigma}(\eps)$ everywhere enters the above
equations multiplied by $V_{\sigma}^2$, the analysis can be recast as the
derivation of shifts in the impurity parameters $\Ed$ and $U$ to account for
a change $\Delta\Gamma_{\sigma}(\eps) = \pi \, \Delta \rho_{\sigma}(\eps) \,
V_{\sigma}^2$ in the spin-dependent hybridization function.

We note that an equation equivalent to Eq.\ \eqref{tE_s} appears in Refs.\
\onlinecite{Martinek:2005} and \onlinecite{Sindel:2007}, which examine the
spin splitting of the impurity level arising from entirely integrating out the
conduction band. In our language, this case corresponds to
$\tilde{\rho}_{\sigma}(\epsilon)=0$ and
$\Delta\rho_{\sigma}(\epsilon)=\rho_{\sigma}(\epsilon)$. These earlier works
did not take into account changes in the energies of the empty and doubly
occupied states that can lead to a shift in the on-site interaction $U$.

\section{Computation of Angular-Momentum Correlations}
\label{app:spin}

This appendix describes the calculation of static correlations
between the impurity spin $\Sd$ defined in Eq.\ \eqref{S_d:def} and one of
$\J_h$, $\J^{m=0}$, and $\J^{m\ne 0}$ representing the total angular
momentum operators of helicity-$h$ electrons, of electrons having orbital
angular momentum $m=0$, and of electrons with angular momentum $m\ne 0$.

Within the numerical renormalization-group treatment of the effective
two-channel Anderson model described by Eq.\ \eqref{H_2chan}, an
appropriate representation of $\J_h$ [defined in Eq.\ \eqref{J_h:def}] for
the calculation of the thermal average $\langle\Sd\cdot\J_h\rangle$ at
temperatures $T\sim D\Lambda^{-N/2}$ is
\begin{equation}
\label{J_h:sum}
\J_h = \sum_{n=0}^N \J_{n,h}
\end{equation}
with
\begin{equation}
\label{J_n,h:def}
\J_{n,h} = \half\sum_{\tau,\tau'} \tf_{n,h,\tau}^{\dag} \:
  \bm{\sigma}_{\tau,\tau'} \: \tf_{n,h,\tau'}^{\pdag},
\end{equation}
where $\tf_{n,h,\tau}$ destroys an electron of total angular momentum $z$
component $\tau=\pm1/2$ on site $n$ of the Wilson chain that
results \cite{Krishna-murthy:1980} from applying the Lanczos procedure to a
discretized version of the bulk Hamiltonian $H_{\bulk}$ in Eq.\
\eqref{H_bulk:diag}.

The calculation of $\langle\Sd\cdot\J^{m=0}\rangle$ and
$\langle\Sd\cdot\J^{m\ne 0}\rangle$, related to $\langle\Sd\cdot\J_h\rangle$
by Eq.\ \eqref{J^m!=0:def}, is more complicated since orbital angular
momentum is not a good quantum number of the bulk states. Since the NRG Lanczos
procedure preserves spin and orbital angular momenta, one can write 
\begin{equation}
\label{f_nht}
\tf_{n,h,\tau}
  = \frac{1}{\sqrt{2}} \bigl( h^{\tau-1/2} f_{n,\tau-1/2,\up}
    + h^{\tau+1/2} f_{n,\tau+1/2,\dn} \bigr)
\end{equation}
in terms of annihilation operators $f_{n,m,\tau}$ for orbital angular momentum
eigenstates, in direct analogy with Eq.\ \eqref{a_kht}.
Substitution of Eq.\ \eqref{f_nht} into Eq.\ \eqref{J_n,h:def} allows one to
write
\begin{equation}
\J_{n,+} + \J_{n,-} = \J_n^{m=0} + \J_n^{m\ne 0} ,
\end{equation}
where
\begin{align}
\label{J_n^m=0:def}
\J_n^{m=0}
&= \half \sum_{\tau,\tau'} f_{n,0,\tau}^{\dag} \:
   \bm{\sigma}_{\tau,\tau'} \: f_{n,0,\tau'}^{\pdag}, \\
\label{J_n^m!=0:def}
\J_n^{m\ne 0}
&= \half \sum_{\tau,\tau'} f_{n,2\tau,-\tau}^{\dag} \:
   \bm{\sigma}_{\tau,\tau'} \: f_{n,2\tau',-\tau'}^{\pdag}, \\
\end{align}
Equation \eqref{J_n^m!=0:def} makes clear that $\J_n^{m\ne 0}$ includes
terms that are off-diagonal in the orbital angular momentum index.

Equation \eqref{f_nht} can be inverted to yield, for $m=0$, $\tau=\pm\half$
and for $m=\pm 1$, $\tau=\mp\half$,
\begin{equation}
\label{f_nmt}
f_{n,m,\tau}
  = \frac{1}{\sqrt{2}} \bigl[\tf_{n,+,\tau} + (-1)^m \tf_{n,-,\tau}\bigr] .
\end{equation}
Substitution of Eq.\ \eqref{f_nmt} into Eqs.\ \eqref{J_n^m=0:def} and
\eqref{J_n^m!=0:def} yields
\begin{align}
\label{J^m=0}
\J^{m=0}    &= \half \sum_{n=0}^N
   \bigl( \J_{n,+} + \J_{n,-} + \tilde{\J}_n \bigr) \\
\label{J^m!=0}
\J^{m\ne 0} &= \half \sum_{n=0}^N
   \bigl( \J_{n,+} + \J_{n,-} - \tilde{\J}_n \bigr) ,
\end{align}
where
\begin{equation}
\label{tJ_n:def}
\tilde{\J}_n = \half \sum_{h,\tau,\tau'} \tf_{n,h,\tau}^{\dag} \:
   \bm{\sigma}_{\tau,\tau'} \: \tf_{n,-h,\tau'}^{\pdag}
\end{equation}
is off-diagonal in the helicity index.

Equations \eqref{J_h:sum}, \eqref{J_n,h:def}, and
\eqref{J^m=0}--\eqref{tJ_n:def} contain the prescription for constructing
the total angular momentum operators in our NRG calculations.
In order to obtain ground-state correlations, we took the thermal
averages of $\Sd$ with $\J_h$, $\J^{m=0}$, and $\J^{m\ne 0}$ in 
the limit of large iteration numbers corresponding to temperature
scales far below $T_K$.


\begin{thebibliography}{99}
\bibitem{units}
Throughout this paper, we work in units where $\hbar=k_B=g_b\mu_B=1$.
\bibitem{shift-note}
We have displaced the angular argument $\phi_k$ by $\pi/2$ to absorb the
prefactor of $i$ in $H_{\Rash}$ as defined in Eq.\ \eqref{H_Rash}.
\end{thebibliography}
\end{document}